\documentclass[aps,prb,twocolumn,floatfix,superscriptaddress,longbibliography]{revtex4-1}
\usepackage{graphicx,txfonts,bm}
\usepackage[colorlinks,citecolor=magenta,linkcolor=blue]{hyperref}

\usepackage{tikz} 
\def\wang{\text{\protect\tikz[baseline=.1ex]{
	\protect\draw[very thick] (-1.6ex,0.6ex)--(1.6ex,0.6ex);
	\protect\draw[very thick] (-1.6ex,0)--(-1.6ex,1.2ex);
	\protect\draw[very thick] (0,0)--(0,1.2ex);
	\protect\draw[very thick] (1.6ex,0)--(1.6ex,1.2ex);
	}}
}

\begin{document}

\title{Stochastic pole expansion method}

\author{Li Huang}
\email{lihuang.dmft@gmail.com}
\affiliation{Science and Technology on Surface Physics and Chemistry Laboratory, P.O. Box 9-35, Jiangyou 621908, China}

\author{Shuang Liang}
\affiliation{Institute of Physics, Chinese Academy of Sciences, Beijing 100190, China}

\date{\today}

\begin{abstract}
In this paper, we propose a new analytic continuation method to extract real frequency spectral functions from imaginary frequency Green's functions of quantum many-body systems. This method is based on the pole representation of Matsubara Green's function and a stochastic sampling procedure is utilized to optimize the amplitudes and locations of poles. In order to capture narrow peaks and sharp band edges in the spectral functions, a constrained sampling algorithm and a self-adaptive sampling algorithm are developed. To demonstrate the usefulness and performance of the new method, we at first apply it to study the spectral functions of representative fermionic and bosonic correlators. Then we employ this method to tackle the analytic continuation problems of matrix-valued Green's functions. The synthetic Green's functions, as well as realistic correlation functions from finite temperature quantum many-body calculations, are used as input. The benchmark results demonstrate that this method is capable of reproducing most of the key characteristics in the spectral functions. The sharp, smooth, and multi-peak features in both low-frequency and high-frequency regions of spectral functions could be accurately resolved, which overcomes one of the main limitations of the traditional maximum entropy method. More importantly, it exhibits excellent robustness with respect to noisy and incomplete input data. The causality of spectral function is always satisfied even in the presence of sizable noises. As a byproduct, this method could derive a fitting formula for the Matsubara data, which provides a compact approximation to the many-body Green's functions. Hence, we expect that this new method could become a pivotal workhorse for numerically analytic continuation and be broadly useful in many applications.
\end{abstract}

\maketitle

\tableofcontents

\section{Introduction\label{sec:intro}}

Matsubara Green's functions $G(i\omega_n)$, or equivalently imaginary time Green's functions $G(\tau)$, of quantum many-body systems are of fundamental importance for finite temperature quantum field theories~\cite{many_body_book,many_body_book_2016}. They are usually generated in finite temperature quantum simulations, such as many-body perturbative calculations~\cite{PhysRev.139.A796,Aryasetiawan_1998,RevModPhys.74.601,PhysRevB.91.235114}, quantum Monte Carlo simulations of impurity, lattice, and condensed matter systems~\cite{RevModPhys.83.349,RevModPhys.73.33,gubernatis_kawashima_werner_2016}, and lattice gauge theory calculations~\cite{PhysRevD.24.2278,PhysRevB.24.4295,ASAKAWA2001459}, just to name a few. In principle, these correlators are not experimentally observable. We have to convert them to retarded Green's functions $G^{R}(\omega)$, or equivalently spectral functions $A(\omega)$, by using analytic continuation. And then the spectra can be compared with the correspondingly spectroscopic data. Clearly, analytic continuation provides a bridge between quantum many-body theories and experimental observations. 

In general the imaginary frequency Green's function $G(i\omega_n)$ and the spectral function $A(\omega)$ are connected by the following Fredholm integral equation of the first kind~\cite{many_body_book}:
\begin{equation}
\label{eq:ac}
G(i\omega_n) = \int^{+\infty}_{-\infty} d\omega~K(\omega_n,\omega) A(\omega),
\end{equation}
where $i$ is the imaginary unit, $\omega_n$ means the Matsubara frequency, and $K(\omega_n,\omega) = 1 / (i\omega_n - \omega)$ is the kernel function. The mapping from $A(\omega)$ to $G(i\omega_n)$ is linear. So, provided $A(\omega)$, it is easy to obtain $G(i\omega_n)$ via numerical integration. However, given $G(i\omega_n)$, seeking a reasonable $A(\omega)$ to satisfy Eq.~(\ref{eq:ac}) needs tremendous efforts. Mathematically, the extraction of $A(\omega)$ is equivalent to carrying out an inverse Laplace transformation, which is a well-known ill-conditioned problem~\cite{PhysRevLett.55.1204,PhysRevB.34.4744}. Because the kernel approaches zero as $\omega$ increases, $A(\omega)$ is very sensitive to the noises embedded in $G(i\omega_n)$. Small fluctuations or noises in $G(i\omega_n)$, which are almost inevitably in quantum many-body calculations~\cite{RevModPhys.73.33,RevModPhys.83.349,gubernatis_kawashima_werner_2016}, could lead to significant changes in $A(\omega)$. Thus, a reliable comparison between theoretical and experimentally observed spectra becomes impossible. This is an intrinsic difficulty of the analytic continuation problems~\cite{PhysRevB.41.2380}, and has long been a key factor limiting the usefulness of finite temperature quantum simulations.

To solve the analytic continuation problems, people have developed many numerical methods in the past decades, including the non-negative least-squares method (NNLS)~\cite{Lawson1995}, non-negative Tikhonov method (NNT)~\cite{Tik1995,PhysRevB.107.085129}, Pad\'{e} approximation (Pad\'{e})~\cite{PhysRevB.61.5147,Vidberg1977,PhysRevB.93.075104,PhysRevB.82.165125,PhysRevD.96.036002,PhysRevB.87.245135}, maximum entropy method (MaxEnt) and its extensions~\cite{JARRELL1996133,PhysRevB.44.6011,PhysRevE.82.026701,PhysRevB.81.155107,PhysRevE.94.023303,PhysRevB.96.155128,PhysRevB.95.121104,PhysRevB.92.060509,PhysRevLett.107.092003,PhysRevB.61.9300,PhysRevB.98.205102,PhysRevB.106.245150}, stochastic analytic continuation (SAC) and its variants~\cite{beach2004,PhysRevB.57.10287,PhysRevE.94.063308,PhysRevE.81.056701,PhysRevB.76.035115,PhysRevX.7.041072,PhysRevB.78.174429}, stochastic optimization method (SOM)~\cite{PhysRevB.62.6317,PhysRevB.95.014102,KRIVENKO2019166,KRIVENKO2022108491,PhysRevB.94.125149}, sparse modeling method (SpM)~\cite{PhysRevE.95.061302,PhysRevB.105.035139}, Nevanlinna analytic continuation (NAC)~\cite{PhysRevLett.126.056402,BNAC}, and Carath\'{e}odory method (Carath\'{e}odory)~\cite{PhysRevB.104.165111}, and so on. In addition, machine learning-assisted analytic continuation methods~\cite{PhysRevLett.124.056401,PhysRevB.98.245101,PhysRevB.105.075112,PhysRevResearch.4.043082,Yao_2022,Arsenault_2017} are also exploited in recent years, but they have not yet been broadly used with realistic quantum Monte Carlo data. 

To our knowledge, perhaps MaxEnt is the most widely used analytic continuation method~\cite{JARRELL1996133,PhysRevB.44.6011}. It dominates this field for quite a long time. In this method, the spectral function is regarded as a probability function and then the Bayesian statistical inference is employed to select the most probable spectrum that maximizes a generalized Shannon-Jaynes entropy~\cite{Bryan1990}. This method is quite efficient, but sometimes it tends to blur the sharp features in the spectrum. Another popular analytic continuation tool is the SAC method~\cite{beach2004,PhysRevB.57.10287,PhysRevE.94.063308}. The spectrum is at first parameterized by a large number of $\delta$ functions in continuous frequency space. Then the amplitudes and locations of the $\delta$ functions are sampled by the Monte Carlo method. Note that various constraints, such as locations of band boundaries or spectral weight of quasiparticle peak, can be encoded in the stochastic sampling procedure. This leads to the constrained SAC method~\cite{PhysRevE.94.063308}. It can resolve intricate spectral functions with both sharp edge features and broad peaks precisely at the cost of computational efficiency~\cite{SHAO20231,PhysRevX.7.041072}. The SOM method is just a cousin of the SAC method~\cite{PhysRevB.62.6317,PhysRevB.95.014102}. The most significant difference is that the SOM method utilizes superposition of many rectangle functions, instead of $\delta$ functions, to parameterize the spectral function. Sometimes both the SOM method and the SAC method are called the average spectrum method (ASM) in the literatures~\cite{PhysRevB.78.174429,PhysRevB.101.085111,PhysRevB.102.035114,doi:10.1063/1.3185728}. Accordingly it is not surprising that the two methods share the same strengths and weaknesses. 

We would like to emphasize that the spirits of the MaxEnt, SAC, and SOM methods are to fit the spectral functions to the imaginary frequency Green's functions. However, there is an alternative route for solving the analytic continuation problems. In the class are methods that aim to interpolate, rather than fit, the imaginary frequency Green's function in the complex plane by using some sorts of rational functions. If the analytic form of the Green's function is established, one may immediately substitute $i\omega_n$ by $\omega + i0^{+}$ to yield the required spectral function. Typical methods in this class include the Pad\'{e}~\cite{Vidberg1977,PhysRevB.93.075104}, NAC~\cite{PhysRevLett.126.056402}, and Carath\'{e}odory~\cite{PhysRevB.104.165111} methods. The Pad\'{e} method requires high-precision input data. Basically, it works well only if the input data on the imaginary axis are not subject to stochastic uncertainty (or noise) and the number of data points is small~\cite{Vidberg1977,PhysRevB.93.075104}. Furthermore, it often generates unphysical oscillation in the high-frequency region of the spectrum, thus violates the causality condition~\cite{PhysRevB.105.035139}. The newly developed NAC method is apparently superior to the Pad\'{e} method. It takes the Nevanlinna analytic structure of the Green's function into account~\cite{PhysRevLett.126.056402}. The evaluated spectral function is guaranteed to be intrinsically positive and normalized. Later, the Nevanlinna interpolation scheme is generalized to treat the matrix-valued Green's functions. This is the so-called Carath\'{e}odory method~\cite{PhysRevB.104.165111}. The two methods can resolve complicated spectral functions over a wide range of frequencies with unprecedented accuracy. However, they are not numerical stable and are not directly applicable to bosonic systems~\cite{BNAC,PhysRevB.107.075151}. More seriously, the two methods are not robust in the presence of noise. If the input data are noisy, the Pick's criterion is violated and the Nevanlinna (Carath\'{e}odory) interpolations won't exist~\cite{PhysRevLett.126.056402}. The obtained spectral functions are not guaranteed to be causal at that time. This deficiency greatly restricts the applications of the two methods in the post-processing procedure for quantum Monte Carlo simulations.

Clearly, though significant progress has been made in solving the inverse problem [see Eq.~(\ref{eq:ac})], it is still far away from being completely settled. This explains why analytic continuation is a long-standing and important problem in computational quantum many-body physics~\cite{many_body_book,many_body_book_2016}. Nevertheless, new analytic continuation methods that based on different principles and strategies are always useful~\cite{PhysRevB.106.245150,PhysRevB.107.085129,PhysRevE.106.025312,PhysRevB.107.075151,LY2022,YING2022111549}. In this paper, we would like to introduce a new analytic continuation method, namely the stochastic pole expansion method (dubbed SPX). It adopts the pole expansion to parameterize the imaginary frequency Green's function. Then the weights and locations of the poles are optimized by stochastic method, hence the name of the method. Finally, the spectral function is evaluated by using the optimal poles. In essence, the SPX method can be classified as the ASM method~\cite{PhysRevB.78.174429}. But it inherits the advantages of both fitting and interpolation approaches. At first, it can recover complicated spectral functions. These spectra usually exhibit some distinctive features, such as large gap, sharp band edge, narrow resonance peak, and long tail etc., over a wide energy range. Second, it provides an approximated pole representation for Matsubara Green's function in the entire complex plane. In other words, a fitting formula for the Green's function is derived once the analytic continuation is finished. Such a formula can serve as a noise filter and a compact representation of many-body Green's function. Third, it is robust with respect to noisy and incomplete data. The sum-rule and causality of the spectral functions are automatically guaranteed. Last but not the least, this method is quite general. It is suitable for not only fermionic but also bosonic correlators. It is straightforward to generalize it to support analytic continuation of matrix-valued Green's functions. Actually, so long as the given correlators can be described by using the Lehmann representation~\cite{many_body_book}, the SPX method always works. 

The rest of this paper is organized as follows. In Section~\ref{sec:formalism}, we at first review the basic properties of the finite temperature Green's function and the Lehmann representation. And then we elaborate on the core idea of the SPX method. The pole representation, and the stochastic approach that is used to optimize the amplitudes and locations of the poles, are explained. In Section~\ref{sec:algorithm}, two auxiliary algorithms, namely the constrained sampling algorithm and the self-adaptive sampling algorithm, are introduced. The remaining part of this section is devoted to implementation details of the SPX method. In the following sections (from Section~\ref{sec:setup} to Section~\ref{sec:application_matrix}), the SPX method is benchmarked thoroughly. The calculated results are compared with those obtained by the traditional MaxEnt method and the exact solutions if available. Firstly, we introduce the computational setups and summarize the test cases (see Section~\ref{sec:setup}). And then the SPX method is utilized to solve the analytic continuation problems for fermionic correlators (see Section~\ref{sec:application_f}), bosonic correlators (see Section~\ref{sec:application_b}), and matrix-valued Green's functions (see Section~\ref{sec:application_matrix}). In Section~\ref{sec:discuss}, we at first focus on the robustness of the SPX method in the presence of sizable noise. The performance of the SPX method is also examined when the input data are incomplete. We analyze the advantages and drawbacks of the SPX method when it is compared to the other analytic continuation methods. Finally, Section~\ref{sec:summary} serves as a short conclusion. We look forward to further applications of the SPX method in other research fields.

\section{Formalisms\label{sec:formalism}}

\subsection{Finite temperature Green's function}

The single-particle imaginary time Green's function $G(\tau)$ reads:
\begin{equation}
G(\tau) = \langle \mathcal{T}_{\tau} c(\tau) c^{\dagger}(0)\rangle,
\end{equation}
where $\mathcal{T}_{\tau}$ is the imaginary time ordering operator, $c^{\dagger}$ and $c$ are creation and annihilation operators in the Heisenberg representation, respectively. For fermions, $G(\tau)$ must fulfil the anti-periodicity condition, i.e., $G(\tau) = -G(\tau + \beta)$. While for bosons, $G(\tau)$ must be $\beta$-periodic, i.e., $G(\tau) = G(\tau + \beta)$. Here $\beta$ denotes the inverse temperature of the system ($\beta = 1/T$). The Matsubara Green's function $G(i\omega_n)$ can be derived from $G(\tau)$ via Fourier transformation,
\begin{equation}
G(i\omega_n) = \int^{\beta}_0 d\tau~e^{-i\omega_n\tau} G(\tau),
\end{equation}
and vice versa,
\begin{equation}
G(\tau) = \frac{1}{\beta} \sum_n e^{i\omega_n\tau} G(i\omega_n).
\end{equation}
Note that $\omega_n = (2n + 1)\pi / \beta$ and $2n\pi / \beta$ for fermions and bosons, respectively, where $n \in \mathbb{Z}$.

\subsection{Spectral representation}

Supposed that the spectral density of the single-particle Green's function is $A(\omega)$, then we have:
\begin{equation}
\label{eq:gtau}
G(\tau) = \int^{+\infty}_{-\infty} d\omega
          \frac{e^{-\tau\omega}}{1 \pm e^{-\beta\omega}}
          A(\omega),
\end{equation}
with the positive (negative) sign for fermionic (bosonic) operators. Similarly, 
\begin{equation}
\label{eq:giw}
G(i\omega_n) = \int^{+\infty}_{-\infty} d\omega
               \frac{1}{i\omega_n - \omega} A(\omega).
\end{equation}
These equations denote the spectral representation of single-particle Green's function. We notice that the SPX method, as well as the other analytic continuation methods that are classified as ASM, are closely related to the spectral representation. Next, we would like to make further discussions about this representation for the fermionic and bosonic correlators.  

\emph{Fermionic correlators}. The spectral density $A(\omega)$ is defined on $(-\infty,\infty)$. It is positive definite, i.e., $A(\omega) \ge 0$. Eq.~(\ref{eq:gtau}) and Eq.~(\ref{eq:giw}) can be reformulated as:
\begin{equation}
G(\tau) = \int^{+\infty}_{-\infty} d\omega~K(\tau,\omega) A(\omega),
\end{equation} 
and
\begin{equation}
\label{eq:spectral_f}
G(i\omega_n) = \int^{+\infty}_{-\infty} d\omega~K(\omega_n,\omega) A(\omega),
\end{equation}
respectively. The kernel functions $K(\tau,\omega)$ and $K(\omega_n,\omega)$ are defined as follows:
\begin{equation}
K(\tau,\omega) = \frac{e^{-\tau\omega}}{1 + e^{-\beta\omega}},
\end{equation}
and
\begin{equation}
\label{eq:kernel_f}
K(\omega_n,\omega) = \frac{1}{i\omega_n - \omega}.
\end{equation}

\emph{Bosonic correlators}. The spectral density $A(\omega)$ obeys the following constraint: $\text{sign}(\omega) A(\omega) \ge 0$. Thus, it is more convenient to define a new function $\tilde{A}(\omega)$ where $\tilde{A}(\omega) = A(\omega)/\omega$. Clearly, $\tilde{A}(\omega)$ is always positive definite. As a result Eq.~(\ref{eq:gtau}) and Eq.~(\ref{eq:giw}) can be rewritten as:
\begin{equation}
G(\tau) = \int^{+\infty}_{-\infty} d\omega~
    K(\tau,\omega)\tilde{A}(\omega),
\end{equation}
and
\begin{equation}
\label{eq:spectral_b}
G(i\omega_n) = \int^{+\infty}_{-\infty} d\omega~
    K(\omega_n,\omega) \tilde{A}(\omega),
\end{equation}
respectively. Now the bosonic kernel $K(\tau,\omega)$ becomes:
\begin{equation}
K(\tau,\omega) = \frac{\omega e^{-\tau\omega}}{1 - e^{-\beta\omega}}.
\end{equation}
Especially, $K(\tau,0) = 1/\beta$. As for $K(\omega_n,\omega)$, its expression is:
\begin{equation}
\label{eq:kernel_b}
K(\omega_n,\omega) = \frac{\omega}{i\omega_n - \omega}.
\end{equation}
Especially, $K(0,0) = -1$. Besides the bosonic Green's function, typical correlator of this kind includes the transverse spin susceptibility $\chi_{+-}(\tau) = \langle S_{+}(\tau) S_{-}(0) \rangle$, where $S_{+} = S_x + iS_y$ and $S_{-} = S_x - i S_y$.

\emph{Bosonic correlators of Hermitian operators}. There is a special case of the previous observable kind with $c = c^{\dagger}$. Here, $A(\omega)$ becomes an odd function, and equivalently, $\tilde{A}(\omega)$ is an even function [i.e., $\tilde{A}(\omega) = \tilde{A}(-\omega)$]. Therefore the limits of integrations in Eq.~(\ref{eq:gtau}) and Eq.~(\ref{eq:giw}) are reduced from $(-\infty,\infty)$ to $(0,\infty)$. So the two equations can be transformed into:
\begin{equation}
G(\tau) = \int^{+\infty}_{0} d\omega~
    K(\tau,\omega)\tilde{A}(\omega),
\end{equation}
and
\begin{equation}
\label{eq:spectral_h}
G(i\omega_n) = \int^{+\infty}_{0} d\omega~
    K(\omega_n,\omega) \tilde{A}(\omega),
\end{equation}
respectively. The corresponding $K(\tau,\omega)$ reads:
\begin{equation}
\label{eq:kernel_h_t}
K(\tau,\omega) = \frac{\omega \left[e^{-\tau\omega} + e^{-(\beta - \tau)\omega}\right]}
                      {1 - e^{-\beta\omega}}.
\end{equation}
Especially, $K(\tau,0) = 2 / \beta$. And $K(\omega_n,\omega)$ becomes:
\begin{equation}
\label{eq:kernel_h}
K(\omega_n, \omega) = \frac{-2\omega^2}{\omega_n^2 + \omega^2}.
\end{equation}
Especially, $K(0,0) = -2$. Perhaps the longitudinal spin susceptibility $\chi_{zz}(\tau) = \langle S_z(\tau) S_z(0) \rangle$ and the charge susceptibility $\chi_{ch}(\tau) = \langle N(\tau) N(0) \rangle$ are the most widely used observables of this kind.

\subsection{Pole representation}

It is well known that the finite temperature many-body Green's functions can be expressed within the Lehmann representation~\cite{many_body_book}:
\begin{equation}
\label{eq:lehmann}
G_{ab}(z) = \frac{1}{Z} \sum_{m,n}
\frac{\langle n | d_a | m \rangle \langle m | d_b^{\dagger} | n \rangle}{z + E_n - E_m}
\left(e^{-\beta E_n} \pm e^{-\beta E_m}\right),
\end{equation}
where $a$ and $b$ are the band indices, $d$ ($d^{\dagger}$) denote the annihilation (creation) operators, $|n \rangle$ and $|m \rangle$ are eigenstates of the Hamiltonian $\hat{H}$, and $E_n$ and $E_m$ are the corresponding eigenvalues, $Z$ is the partition function ($Z = \sum_n e^{-\beta E_n}$). The positive sign corresponds to fermions, while the negative sign corresponds to bosons. The domain of this function is on the complex plane, but the real axis is excluded ($z \in \{0\} \bigcup \mathbb{C}~\backslash~\mathbb{R} $). If $z = i\omega_n \in i\mathbb{R}$, $G_{ab}(i\omega_n)$ is the Matsubara Green's function. If $z = \omega + i0^{+}$, $G_{ab}(\omega + i0^{+}) = G_{ab}^{R}(\omega)$ is called the retarded Green's function.

At first, we focus on the diagonal cases ($a = b$). For the sake of simplicity, the band indices are ignored in the following discussions. Let $A_{mn} = \langle n | d | m \rangle \langle m | d^{\dagger} | n \rangle \left(e^{-\beta E_n} + e^{-\beta E_m}\right) / Z$ and $P_{mn} = E_m - E_n$, then $G(z) = \sum_{m,n} A_{mn} / (z - P_{mn})$~\cite{PhysRevB.107.075151}. Clearly, only those nonzero elements of $A_{mn}$ contribute to the Green's function. If the indices $m$ and $n$ are further compressed into $\gamma$ (i.e, $\gamma = \{m,n\}$), then Eq.~(\ref{eq:lehmann}) is simplified to:
\begin{equation}
\label{eq:pole}
G(z) = \sum^{N_p}_{\gamma = 1} \frac{A_{\gamma}}{z - P_{\gamma}}.
\end{equation}
Here, $A_{\gamma}$ and $P_{\gamma}$ mean the amplitude and location of the $\gamma$-th pole, respectively. $N_p$ means the number of poles, which is equal to the total number of nonzero $A_{mn}$. Such an analytic expression of Green's function is called the \emph{pole expansion}. It is valid for both fermionic and bosonic correlators.

\emph{Fermionic correlators}. For fermionic systems, the pole representation for Matsubara Green's function can be recast as:
\begin{equation}
\label{eq:pole_f}
G(i\omega_n) = \sum^{N_p}_{\gamma = 1} \Xi(\omega_n, P_{\gamma}) A_{\gamma}.
\end{equation}
Here, $\Xi$ is called the kernel matrix. It is evaluated by:
\begin{equation}
\label{eq:xi_f}
\Xi(\omega_n, \omega) = \frac{1}{i\omega_n - \omega}.
\end{equation}
Note that $A_{\gamma}$ and $P_{\gamma}$ should satisfy the following constraints:
\begin{equation}
\label{eq:sum_rule_f}
\forall \gamma,~0 \le A_{\gamma} \le 1,~\sum_{\gamma} A_{\gamma} = 1,~P_{\gamma} \in \mathbb{R}.
\end{equation}

\emph{Bosonic correlators}. For bosonic systems, the pole representation for Matsubara Green's function can be defined as follows:
\begin{equation}
\label{eq:pole_b}
G(i\omega_n) = \sum^{N_p}_{\gamma=1} \Xi(\omega_n, P_{\gamma}) \tilde{A}_{\gamma}.
\end{equation}
Here, $\Xi$ is evaluated by:
\begin{equation}
\label{eq:xi_b}
\Xi(\omega_n, \omega) = \frac{G_0 \omega}{i\omega_n - \omega},
\end{equation}
where $G_{0} = -G(i\omega_n = 0)$, which should be a positive real number. Be careful, $\Xi(0,\omega) = -G_0$. $\tilde{A}_{\gamma}$ is the renormalized amplitude of the $\gamma$-th pole:
\begin{equation}
\tilde{A}_{\gamma} = \frac{A_{\gamma}}{G_0 P_{\gamma}}.
\end{equation}
Note that $\tilde{A}_{\gamma}$ and $P_{\gamma}$ should satisfy the following constraints:
\begin{equation}
\label{eq:sum_rule_b}
\forall \gamma,~0 \le \tilde{A}_{\gamma} \le 1,~\sum_{\gamma} \tilde{A}_{\gamma} = 1,~P_{\gamma} \in \mathbb{R}.
\end{equation}

\emph{Bosonic correlators of Hermitian operators}. Its pole representation can be defined as follows ($\forall \gamma$, $A_{\gamma} > 0$ and $P_{\gamma} > 0$):
\begin{eqnarray}
\label{eq:pole_h}
G(i\omega_n) &=& \sum^{N_p}_{\gamma = 1}
\left(
    \frac{A_{\gamma}}{i\omega_n - P_{\gamma}} -
    \frac{A_{\gamma}}{i\omega_n + P_{\gamma}}
\right) \nonumber \\
&=& \sum^{N_p}_{\gamma = 1}
\Xi(\omega_n, P_{\gamma}) \tilde{A}_{\gamma}.
\end{eqnarray}
Thus, the kernel matrix $\Xi$ reads:
\begin{equation}
\label{eq:xi_h}
\Xi(\omega_n, \omega) = \frac{-G_0 \omega^2}{\omega^2_n + \omega^2}.
\end{equation}
Especially, $\Xi(0,0) = -G_0$. The renormalized weight $\tilde{A}_{\gamma}$ reads:
\begin{equation}
\tilde{A}_{\gamma} = \frac{2A_{\gamma}}{G_0 P_{\gamma}}.
\end{equation}
The constraints for $\tilde{A}_{\gamma}$ and $P_{\gamma}$ are the same with Eq.~(\ref{eq:sum_rule_b}).

As for the off-diagonal cases ($a \neq b$), it is lightly to prove that $\sum_{\gamma} A_{\gamma} = 0$. It implies that there exist poles with negative weights. Hence we can split the poles into two groups according to the signs of their amplitudes. The Matsubara Green's function can be expressed as follows:
\begin{eqnarray}
\label{eq:pole_m}
G(i\omega_n) &=& \sum^{N^{+}_p}_{\gamma = 1} 
               \frac{A^{+}_{\gamma}}{i\omega_n - P^{+}_{\gamma}} -
               \sum^{N^{-}_p}_{\gamma = 1}
               \frac{A^{-}_{\gamma}}{i\omega_n - P^{-}_{\gamma}} \nonumber \\
             &=& \sum^{N^{+}_p}_{\gamma = 1}
               \Xi(\omega_n, P^{+}_{\gamma}) A^{+}_{\gamma} -
               \sum^{N^{-}_p}_{\gamma = 1}
               \Xi(\omega_n, P^{-}_{\gamma}) A^{-}_{\gamma}
.
\end{eqnarray}
Here, $\Xi(\omega_n, \omega)$ is already defined in Eq.~(\ref{eq:xi_f}). The $A^{\pm}_{\gamma}$ and $P^{\pm}_{\gamma}$ are restricted by Eq.~(\ref{eq:sum_rule_f}). In addition,
\begin{equation}
N_p = N^{+}_p + N^{-}_p,
\end{equation}
and
\begin{equation}
\label{eq:sum_rule_m}
\sum^{N^{+}_p}_{\gamma = 1} A^{+}_{\gamma} - 
\sum^{N^{-}_p}_{\gamma = 1} A^{-}_{\gamma} = 0.
\end{equation}

\subsection{Stochastic optimization}

\begin{figure}[t]
\includegraphics[width=\columnwidth]{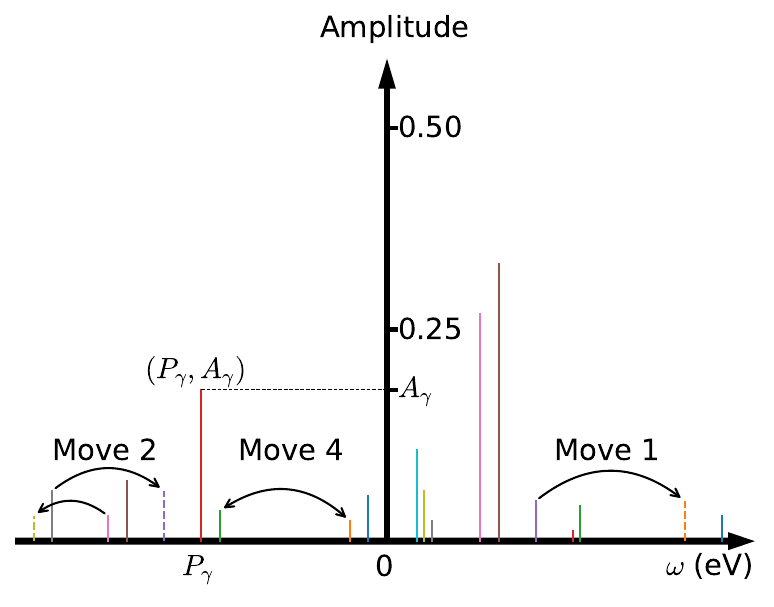}
\caption{Schematic picture for the pole representation of the Matsubara Green's function. The poles are visualized by vertical colorful bars. ``Move 1'', ``Move 2'', and ``Move 4'' denote three possible Monte Carlo updates: (i)~shift a randomly selected pole, (ii) shift two randomly selected poles, and (iii)~swap two randomly selected poles. See the main text for more details. \label{fig:MC}}
\end{figure}

Supposed that the input Matsubara Green's function is $\mathcal{G}(i\omega_n)$, where $n =$ 1, 2, $\cdots$, $N_{\omega}$, the objective of analytic continuation is to fit the (possibly noisy and incomplete) Matsubara data into Eq.~(\ref{eq:pole}) under some constraints. In mathematical language, we should solve the following multivariate optimization problem:
\begin{equation}
\label{eq:chi2}
\mathop{\arg\min}\limits_{ \left\{A_{\gamma}, P_{\gamma}\right\}^{N_p}_{\gamma = 1} } \chi^{2}\left[\left\{A_{\gamma}, P_{\gamma}\right\}^{N_p}_{\gamma = 1}\right].
\end{equation}
Here, $\chi^{2}\left[\left\{A_{\gamma}, P_{\gamma}\right\}^{N_p}_{\gamma = 1}\right]$ is the so-called goodness-of-fit function or loss function. Its definition is as follows:
\begin{equation}
\chi^{2}
\left[\left\{A_{\gamma}, P_{\gamma}\right\}^{N_p}_{\gamma = 1}\right] 
= \frac{1}{N_{\omega}}\sum^{N_{\omega}}_{n = 1}
\left|\left|
\mathcal{G}(i\omega_n) - \sum^{N_p}_{\gamma = 1} \frac{A_{\gamma}}{i\omega_n - P_{\gamma}} 
\right|\right|^2_{F},
\end{equation}
where $||\cdot||_{F}$ denotes the Frobenius norm. The minimization of Eq.~(\ref{eq:chi2}) is highly non-convex. Traditional gradient-based optimization methods, such as the non-negative least squares method, conjugate gradient method, Newton and quasi-Newton methods, are frequently trapped in local minima~\cite{optimization_book}. Their optimized results strongly depend on the initial guess. The semi-definite relaxation (SDR) fitting method~\cite{PhysRevB.101.035143,PhysRevB.107.075151}, adaptive Antoulas-Anderson (AAA) algorithm~\cite{AAA_Lawson,AAA}, and conformal mapping plus Prony's method~\cite{LY2022,YING2022111549}, which have been employed to search the locations of poles in previous works~\cite{PhysRevB.107.075151}, are also tested. But these methods usually fail when $N_p$ is huge [$N_p \sim O(10^3)$] or the Matsubara data are noisy.

In order to overcome the above obstacles, we employ the simulated annealing method~\cite{SA1983} to locate the global minimum of $\chi^{2}$. The core idea is as follows: First of all, a set of $\{A_{\gamma}, P_{\gamma}\}$ parameters are generated randomly. These parameters form a configuration space $\mathcal{C} = \{A_{\gamma}, P_{\gamma}\}$. Second, this configuration space is sampled by using the Metropolis Monte Carlo algorithm. In the present SPX method, four Monte Carlo updates are supported (see Figure~\ref{fig:MC}). They include: (i) Select one pole randomly and shift its location. (ii) Select two poles randomly and shift their locations. (iii) Select two poles randomly and change their amplitudes. The sum-rules, such as Eq.~(\ref{eq:sum_rule_f}) and Eq.~(\ref{eq:sum_rule_b}), should be respected in this update. (iv) Select two poles randomly and exchange their amplitudes. Assumed that the current Monte Carlo configuration is $\mathcal{C} = \{A_{\gamma}, P_{\gamma}\}$, the new one is $\mathcal{C}' = \{A'_{\gamma}, P'_{\gamma}\}$, and $\Delta \chi^2 = \chi^2(\mathcal{C}') - \chi^2(\mathcal{C})$, then the transition probability reads:
\begin{equation}
\label{eq:prob}
p(\mathcal{C} \to \mathcal{C}') =
\left\{
    \begin{array}{lr}
        \exp\left(-\frac{\Delta \chi^2}{2\Theta}\right), & \text{if}~\Delta \chi^2 > 0, \\
        1.0, & \text{if}~\Delta \chi^2 \le 0,
    \end{array}
\right.
\end{equation}
where $\Theta$ is an artificial system temperature and $\chi^2$ is interpreted as energy of the system. Third, the above two steps should be restarted periodically to avoid being trapped by local minima. Fourth, once all the Monte Carlo sampling tasks are finished, we should pick up the ``best'' solution which exhibits the smallest $\chi^2$, or select some ``good'' solutions with small $\chi^2$ and evaluate their arithmetic average. Finally, with the optimized $N_{p}$, and $A_{\gamma}$, and $P_{\gamma}$ parameters, the retarded Green's function $G^{R}(\omega)$ can be easily evaluated by replacing $i\omega_n$ with $\omega + i\eta$ in Eqs.~(\ref{eq:pole_f}),~(\ref{eq:pole_b}),~(\ref{eq:pole_h}), and~(\ref{eq:pole_m}), where $\eta$ is a positive infinitesimal number. And the spectral density $A(\omega)$ is calculated by:
\begin{equation}
A(\omega) = - \frac{1}{\pi} \text{Im} G^{R}(\omega).
\end{equation}

\section{Basic algorithms\label{sec:algorithm}}

\subsection{Constrained sampling algorithm}

In the SPX method, the amplitudes and locations of the poles should be optimized by the Monte Carlo algorithm under some constraints [see Eqs.~(\ref{eq:sum_rule_f}), (\ref{eq:sum_rule_b}), and (\ref{eq:sum_rule_m})]. We note that these constraints are from the canonical relations for the fermionic and bosonic operators~\cite{many_body_book,many_body_book_2016}. They must be satisfied, or else the causality of the spectrum can not be guaranteed. But beyond that, more constraints are allowable. Further restrictions on the amplitudes and locations of the poles can greatly reduce the configuration space that needs to be sampled and enhance the possibility to reach the global minimum of the optimization problem. The possible strategies include: (1) Restrict $\{A_{\gamma}\}$ only; (2) Restrict $\{P_{\gamma}\}$ only; and (3) Restrict both $\{A_{\gamma}\}$ and $\{P_{\gamma}\}$ at the same time. These extra constraints can be deduced from \emph{a priori} knowledge about the Matsubara Green's function $G(i\omega_n)$ and the spectral density $A(\omega)$. For example, for a molecule system, the amplitudes of the poles are likely close. On the other hand, if we know nothing about the input data and the spectra, we can always try some constraints. The universal trend is that the more reasonable the constraints, the smaller the $\chi^2$ function. This is the so-called \emph{constrained sampling algorithm}. By combining it with the SPX method (dubbed C-SPX), the ability to resolve fine features in the spectra will be greatly enhanced. To the best of our knowledge, the constrained sampling algorithm was first proposed by A. W. Sandvik~\cite{PhysRevE.94.063308}. And then it is broadly used in analytic continuations for spin susceptibilities of quantum many-body systems~\cite{PhysRevX.7.041072,PhysRevB.98.174421}. Quite recently, Shao and Sandvik summarized various approaches to mount the constraints and benchmark their performances in a comprehensive review concerning the SAC method~\cite{SHAO20231}. Due to the similarities of the SPX and SAC methods, it is believed that all the constraint tactics as suggested in Reference~[\onlinecite{SHAO20231}] should be applicable for the SPX method.
 
\subsection{Self-adaptive sampling algorithm}

In analogy to the SAC method, the poles in the SPX method are distributed randomly in a real frequency grid~\cite{beach2004,PhysRevB.57.10287}. This grid must be extremely dense and is usually linear. But in principle a nonuniform grid is possible. For example, Shao and Sandvik~\cite{SHAO20231} have suggested a nonlinear grid with monotonically increasing spacing for the $\delta$ functions which are used to parameterize a spectrum that exhibits a sharp band edge. Since a spectral density can be viewed as a probability distribution~\cite{JARRELL1996133} and we notice that the distribution of the poles looks quite similar to the spectrum. So, it is natural to adjust the frequency grid dynamically to make sure that the grid density has an approximate distribution with the spectral density as obtained in the previous run. We adopt the following algorithm to manipulate the desirable frequency grid: (1) Calculate the integrated spectral function $\phi(\epsilon)$:
\begin{equation}
\label{eq:tos}
\phi(\epsilon) = \int^{\epsilon}_{\omega_{\text{min}}} A(\omega) d\omega,~\epsilon \in [\omega_{\text{min}},\omega_{\text{max}}].
\end{equation}
(2) The new frequency grid $f_i$ is evaluated by:
\begin{equation}
\label{eq:new_mesh}
f_i = \phi^{-1}(\lambda_i),~i = 1, \cdots, N_f, 
\end{equation}
where $\lambda_i$ is a linear mesh in $[\phi(\omega_{\text{min}}),\phi(\omega_{\text{max}})$], and $N_f$ denotes the number of grid points. Next, we should perform the analytic continuation simulation again to yield a new spectrum, then repeat steps (1) and (2). We find that the $\chi^2$ drops quickly in the first few iterations, and then approaches a constant value slowly. The spectrum is refined simultaneously. During the iterations, the frequency grid for the poles is adaptively refreshed via Eqs.~(\ref{eq:tos}) and (\ref{eq:new_mesh}), thus we call it the \emph{self-adaptive sampling algorithm}. It is actually a new variation of the constrained sampling algorithm~\cite{PhysRevE.94.063308}. More importantly, it is quite effective. According to our experiences, 3 $\sim$ 5 iterations are enough to obtain a convergent solution. In practice, we often adopt the spectrum generated by the MaxEnt method to initialize the frequency grid, and then employ the SPX method (dubbed SA-SPX) to refine this spectrum further.      

\subsection{Reference implementation}

\begin{figure*}[ht]
\centering
\includegraphics[width=\textwidth]{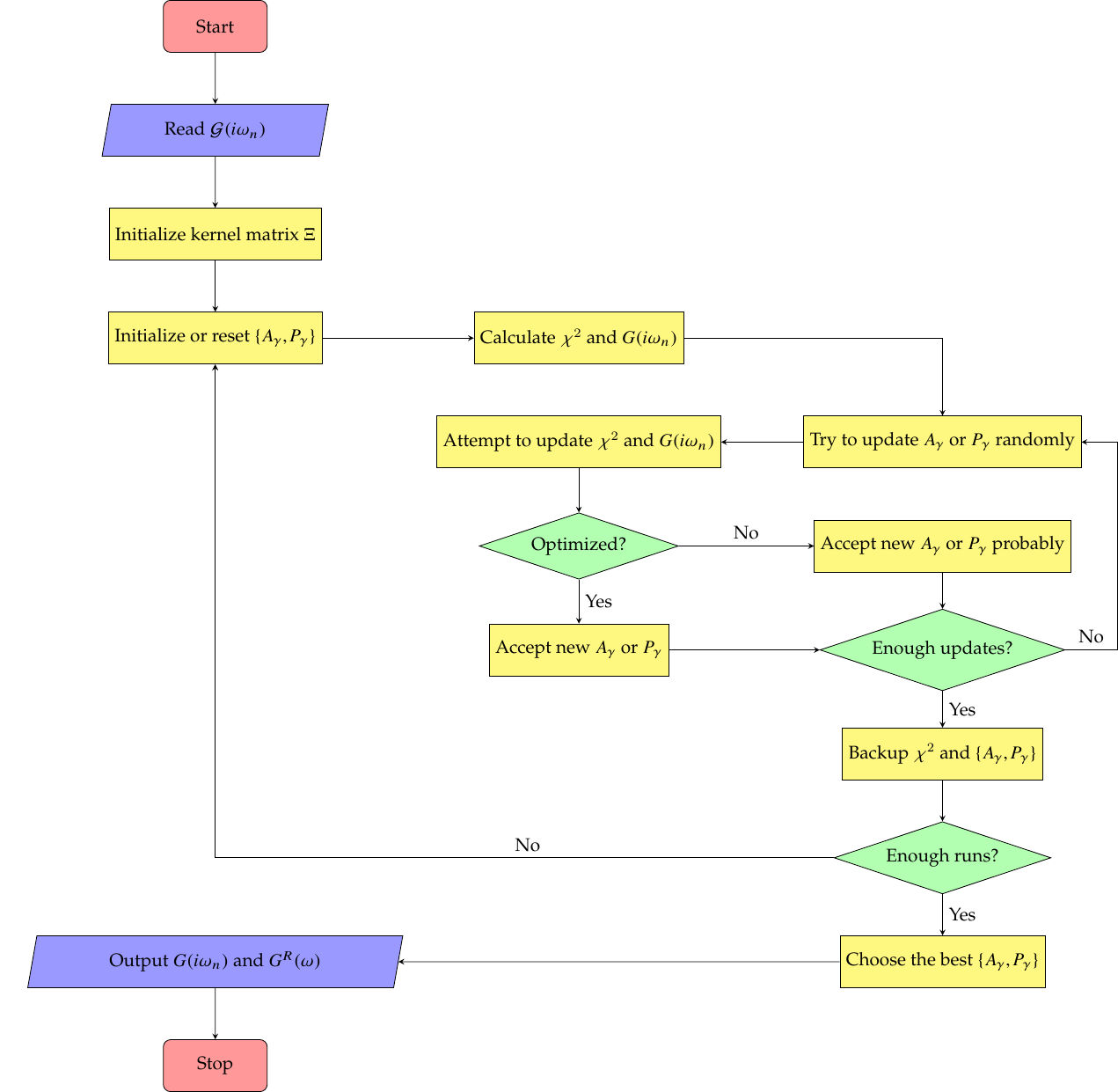}
\caption{Workflow of the SPX method as implemented in the \texttt{ACFlow} code~\cite{Huang:2022}. The $\mathcal{G}(i\omega_n)$, $G(i\omega_n)$, and $G^{R}(\omega)$ mean the input Matsubara Green's function, reconstructed Green's function, and retarded Green's function, respectively. For bosonic functions, $A_{\gamma}$ is replaced by $\tilde{A}_{\gamma}$. \label{fig:diag}}
\end{figure*}

The SPX method, together with the MaxEnt method~\cite{JARRELL1996133,PhysRevB.44.6011}, have been implemented in an open source software package, namely \texttt{ACFlow}~\cite{Huang:2022}, which is a full-fledged analytic continuation toolkit. This package supports analytic continuation for both fermionic and bosonic correlators.

The flowchart of the SPX method is illustrated in Figure~\ref{fig:diag}. Next, we would like to explain some essential steps. (1) \emph{Initialize the kernel matrix $\Xi$}. It is defined in Eqs.~(\ref{eq:xi_f}), (\ref{eq:xi_b}), and (\ref{eq:xi_h}). We should evaluate and save them during the initialization stage to improve the computational efficiency. (2) \emph{Initialize or reset $\{A_{\gamma},P_{\gamma}\}$}. For each Monte Carlo run, $\{A_{\gamma},P_{\gamma}\}$ must be reinitialized to avoid getting trapped in local minima. (3) \emph{Calculate $\chi^{2}$ and $G(i\omega_n)$}. Here, $G(i\omega_n)$ means the reproduced Matsubara data. It can be derived by using the present $\{A_{\gamma},P_{\gamma}\}$ via Eqs.~(\ref{eq:pole_f}), (\ref{eq:pole_b}), (\ref{eq:pole_h}), and (\ref{eq:pole_m}). The loss function $\chi^{2}$ measures the distance between the input Matsubara data $\mathcal{G}(i\omega_n)$ and the reproduced Matsubara data $G(i\omega_n)$. It is evaluated by Eq.~(\ref{eq:chi2}). (4) \emph{Try to update $A_{\gamma}$ or $P_{\gamma}$ randomly}. Four Monte Carlo updates (Move 1 $\sim$ Move 4) are introduced to change the randomly chosen $A_{\gamma}$ or $P_{\gamma}$. (5) \emph{Accept new $A_{\gamma}$ or $P_{\gamma}$}. When $\Delta \chi^{2} < 0$, the Monte Carlo proposal is always accepted. (6) \emph{Accept new $A_{\gamma}$ or $P_{\gamma}$ probably}. When $\Delta \chi^{2} > 0$, it is still possible to accept the Monte Carlo proposal. The transition probability is determined by Eq.~(\ref{eq:prob}). In a standard simulated annealing algorithm, $\Theta$ should decline gradually from high temperature to low temperature~\cite{Marinari:1996dh}. However, in the present implementation, we just set $\Theta$ to a large value (about $10^{6} \sim 10^{9}$). We find that such a large $\Theta$ is essential to enable the Monte Carlo walker to escape the local minima and visit as many configurations as possible. (7) \emph{Backup $\chi^2$ and $\{A_{\gamma},P_{\gamma}\}$}. For each Monte Carlo run, we always record the best solution, i.e., the smallest $\chi^{2}$ and the corresponding $\{A_{\gamma},P_{\gamma}\}$. (8) \emph{Enough runs}? The Monte Carlo procedure is repeated many times in order to find out the ``true'' global minimum of $\chi^{2}$. (9) \emph{Choose the best $\{A_{\gamma},P_{\gamma}\}$}. Once the outer iteration is finished, we obtain a collection of $\chi^{2}$ and $\{A_{\gamma},P_{\gamma}\}$. For the molecule cases, we should pick up the smallest $\chi^2$ and the corresponding $\{A_{\gamma},P_{\gamma}\}$. However, for the condensed matter cases, it would be better to select some ``good'' solutions and evaluate their averaged value. (10) \emph{Output $G(i\omega_n)$ and $G^{R}(\omega)$}. Finally, the Matsubara Green's function $G(i\omega_n)$ and retarded Green's function $G^{R}(\omega)$ are calculated via Eq.~(\ref{eq:pole}) using the optimized $\{A_{\gamma},P_{\gamma}\}$.   

In the present implementation, several numerical issues need to be emphasized.
 
\begin{itemize}
\item The kernel matrix $\Xi(\omega_n,\omega)$ should be computed at advance and kept in memory. This allows us to calculate the transition probability $p$ and the Green's function $G(i\omega_n)$ efficiently. The Matsubara frequencies $\omega_n$ are taken from the input data directly. And we have to create a frequency grid on the real axis for $\omega$ ($\omega \in [\omega_{\text{min}},\omega_{\text{max}}]$). In principle, the poles are distributed in a continuous frequency space. So in order to improve the computational accuracy, the number of grid points ($N_f$) should be as large as possible.

\item According to Eq.~(\ref{eq:prob}), even when $\chi^2(\mathcal{C}') > \chi^2(\mathcal{C})$, the transition from $\mathcal{C}'$ to $\mathcal{C}$ is not always rejected. Though the transition probability $p$ could be quite small, we still have the chance to accept a less optimal solution than what we currently have and escape the local minima. This transition probability is largely controlled by the $\Theta$ parameter, which behaves as the system's annealing temperature~\cite{SA1983,Marinari:1996dh}. In accordance with the spirit of the simulated annealing algorithm, $\Theta$ should be decreased gradually. However, in the present implementation, $\Theta$ is fixed and considered as a user-supplied parameter. According to our experiences, the preferred value of $\Theta$ is between $10^{6}$ and $10^{9}$.

\item In the SPX method, the Monte Carlo sampling procedure should be repeated from scratch many times. In each run, the $\chi^{2}$ and the corresponding Monte Carlo configuration $\mathcal{C}$ will be tracked. We find that when the final spectral density exhibits multiple $\delta$-like peaks, it is wise to pick up the ``best'' solution whose $\chi^{2}$ is the smallest. However, when the spectral density is expected to be broad and smooth, it is better to calculate an arithmetic average of some selected ``good'' solutions. We just adopted the following strategy to filter the solutions~\cite{KRIVENKO2019166,KRIVENKO2022108491}. At first, we try to calculate the median or mean value of the collected $\chi^{2}$ data, i.e., $\langle \chi^2 \rangle$. Then, only the solutions, whose $\chi^{2}$ are smaller than $\langle \chi^2 \rangle / \alpha_{\text{good}}$, are retained. Here, $\alpha_{\text{good}}$ is an adjustable parameter. Its optimal value is around $1.0 \sim 1.2$.

\item The SPX method has been parallelized to improve computational efficiency. It is straightforward to launch multiple Monte Carlo processes simultaneously to accelerate the calculation.
\end{itemize}

\section{Computational setup\label{sec:setup}}

\begin{table*}[ht]
\centering
\begin{tabular}{llll}
\hline
\hline
System & Model & Features & Sections \\
\hline
~ & M$_{01}$: Gaussian model & Multiple broad peaks + sharp quasiparticle peak & \ref{subsec:gauss_f} and \ref{subsec:noise} \\
~ & M$_{02}$: Pole model & Multiple off-centered $\delta$ peaks & \ref{subsec:pole_f} and \ref{subsec:noise} \\
Fermionic & M$_{03}$: Resonance model & Sharp band edges + big gap + wide platform & \ref{subsec:resonance_f} \\
correlators & M$_{04}$: Matsubara Green's function & Sharp quasiparticle peak + lower and upper Hubbard bands & \ref{subsec:green_f} \\
~ & M$_{05}$: Matsubara self-energy function & - & \ref{subsec:self_f} \\
\hline
~ & M$_{06}$: Gaussian model & Two broad peaks & \ref{subsec:gauss_b} \\
~ & M$_{07}$: Pole model & Two off-centered $\delta$ peaks & \ref{subsec:pole_b} \\
Bosonic & M$_{08}$: Resonance model & Sharp band edge + wide platform & \ref{subsec:resonance_b} \\
correlators & M$_{09}$: Spin-spin correlation function & Sharp band edges + quasilinear behavior & \ref{subsec:chi_b} \\
~ & M$_{10}$: Current-current correlation function & Narrow Drude peak + broad interband transition peak & \ref{subsec:sigma_b} and \ref{subsec:incomplete} \\
~ & M$_{11}$: Lindhard function & - & \ref{subsec:lindhard_b} \\
\hline
Matrix-valued & M$_{12}$: Gaussian model & Multiple broad peaks & \ref{subsec:gauss_m} \\
correlators & M$_{13}$: Pole model & Multiple off-centered $\delta$ peaks & \ref{subsec:pole_m} \\
\hline
\hline
\end{tabular}
\caption{Overview of the test cases. The matrix-valued correlators are fermionic. Note that only the M$_{04}$, M$_{05}$, and M$_{09}$ cases are taken from realistic quantum many-body simulations. The other cases are designed to represent typical spectra one would encounter in practice. \label{tab:examples}}
\end{table*}

To benchmark the SPX method, 13 test examples (namely M$_{01}$ $\sim$ M$_{13}$), including the fermionic correlators, bosonic correlators, and matrix-valued Green's functions, are established. The spectral functions are representative, as one would encounter in practice. An overview of these examples is presented in Table~\ref{tab:examples}, and their details will be explicitly described in the following sections. All the analytic continuation calculations were done by using the \texttt{ACFlow} toolkit~\cite{Huang:2022}. 

We mainly concentrate on test functions that are known in the entire complex plane. At first, the test functions are evaluated at the Matsubara frequency axis. Since the realistic Matsubara data from finite temperature quantum Monte Carlo simulations are usually noisy~\cite{gubernatis_kawashima_werner_2016,RevModPhys.83.349,RevModPhys.73.33}, multiplicative Gaussian noise will be manually added to the clean Matsubara data to mimic this situation. We adopted the following formula~\cite{PhysRevB.107.075151}:
\begin{equation}
\label{eq:noise}
\mathcal{G}_{\text{noisy}} = \mathcal{G}_{\text{exact}}[1 + \delta N_{\mathbb{C}}(0,1)],
\end{equation}
where $N_{\mathbb{C}}(0,1)$ is the complex-valued normal Gaussian noise, and $\delta$ denotes the noise level of the data. Except stated explicitly, $\delta = 10^{-4}$, the size of input data is $N_{\omega} = 10$, and the standard deviations of these data are fixed to $10^{-4}$.     

Then the Matsubara data are analytically continued to the real axis and compared with the exact solutions. The size of the real frequency grid for computing $\Xi$ is fixed to $N_f = 10^{5}$. If the spectral density manifests broad and smooth peaks, the number of poles is 2000, $\Theta = 10^8$, $\eta = 10^{-3}$, the number of individual Monte Carlo runs is 2000 and each run contains $2 \times 10^5$ Monte Carlo sampling steps. If the spectral density exhibits multiple $\delta$-like peaks, the number of poles is considered as \emph{a priori} knowledge, $\Theta = 10^{6}$, $\eta = 10^{-2}$, the number of individual Monte Carlo runs is 1000 and each one contains $4 \times 10^4 \times number~of~poles$ Monte Carlo sampling steps.
 
The traditional MaxEnt method is also employed~\cite{JARRELL1996133,PhysRevB.44.6011} so as to crosscheck the analytic continuation results. The ``$\chi^{2}$kink'' algorithm~\cite{PhysRevE.94.023303} is used to search the optimal regulation parameter $\alpha$. The maximum value of $\alpha$ is $10^{9} \sim 10^{15}$, and the number of $\alpha$ parameters are $12 \sim 20$. The default model is flat. The kernel functions are evaluated by Eqs.~(\ref{eq:kernel_f}), (\ref{eq:kernel_b}), and (\ref{eq:kernel_h}).

\section{Applications: Fermionic correlators\label{sec:application_f}}

In this section, the SPX method is employed to extract spectral functions from various fermionic correlators.

\subsection{Gaussian model\label{subsec:gauss_f}}

\begin{figure*}[ht]
\includegraphics[width=\textwidth]{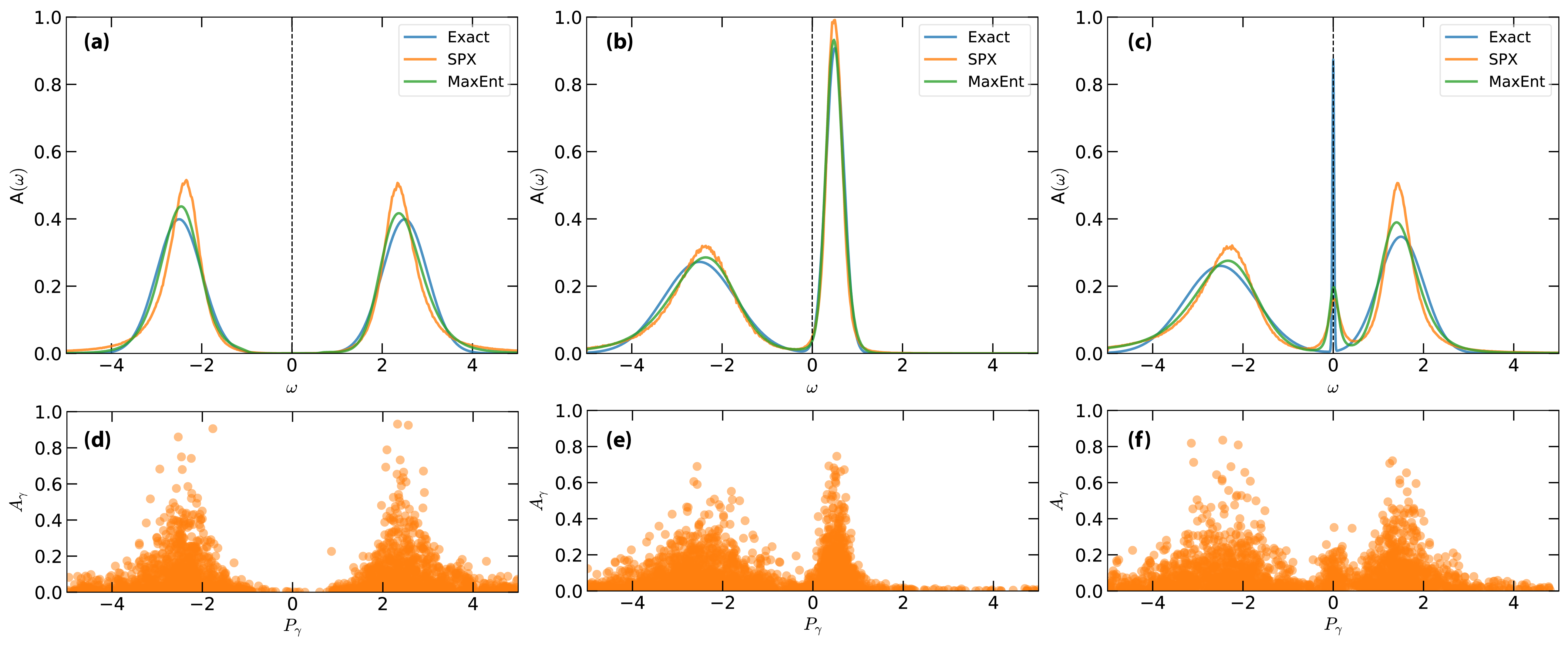}
\caption{Analytic continuations of fermionic correlators (M$_{01}$: Gaussian model). (a)-(c) Exact and calculated spectral functions. The vertical dashed lines denote the Fermi level. (d)-(f) Typical distributions of poles that corresponding to the ``best'' solutions. The amplitudes of poles presented in the lower panels have been rescaled for a better view. \label{fig:X01}}
\end{figure*}

We at first benchmark the SPX method on the condensed matter cases, in which the spectral functions are usually broad and smooth. The spectral functions are assumed to be the superposition of a few Gaussian peaks: 
\begin{equation}
\label{eq:model01}
A(\omega) = \sum^{N_g}_{i=1} w_i \exp \left[\frac{-(\omega - \epsilon_i)^2}{2 \Gamma^2_i}\right], 
\end{equation}
where $N_g$ is the number of Gaussian peaks, $w_i$, $\epsilon_i$, and $\Gamma_i$ denote the weight, position, and broadening of the $i$-th Gaussian peak, respectively. The spectral functions are then back-continued to the Matsubara frequency axis by using Eq.~(\ref{eq:spectral_f}) with $\beta = 10.0$. The Matsubara data, supplemented with random noises, are used as inputs for the SPX method.

Figure~\ref{fig:X01} shows the analytic continuation results by the SPX method for three typical scenarios: (i) two broad Gaussian peaks separated by a sizable gap ($\Delta_{\text{gap}} \approx 2.0$, $N_g = 2$, $w_1 = w_2 = 0.5$, $\epsilon_1 = -\epsilon_2 = 2.5$, $\Gamma_1 = \Gamma_2 = 0.5$), (ii) two Gaussian peaks with a ``pseudogap'' ($N_g = 2$, $w_1 = 1.0$, $w_2 = 0.3$, $\epsilon_1 = 0.5$, $\epsilon_2 = -2.5$, $\Gamma_1 = 0.2$, $\Gamma_2 = 0.8$), and (iii) two Gaussian peaks plus a sharp quasiparticle resonance peak ($N_g = 3$, $w_1 = 1.0$, $w_2 = 0.3$, $w_3 = 0.4$, $\epsilon_1 = 0.0$, $\epsilon_2 = -2.5$, $\epsilon_3 = 1.5$, $\Gamma_1 = 0.02$, $\Gamma_2 = 0.8$, $\Gamma_3 = 0.5$). As is evident in Fig.~\ref{fig:X01}(a)-(c), the major features of the synthetic spectral functions, including the energy gap, positions and weights of the peaks, are well recovered by the SPX method. The only exception is that the full width at half maximum and height of the quasiparticle resonance peak for scenario (iii) [see Fig.~\ref{fig:X01}(c)] are somewhat overestimated. The MaxEnt method~\cite{JARRELL1996133,PhysRevB.44.6011} leads to slightly better results for scenarios (i) and (ii). But it is also unable to recover the sharp quasiparticle resonance peak for scenario (iii). Fig.~\ref{fig:X01}(d)-(f) illustrate the distributions of poles for the three scenarios. Not surprisingly, they manifest approximate characteristics to the correspondingly spectral functions. Overall, for the condensed matter cases, the performance of the SPX method is comparable with the other fitting-based analytic continuation methods.         

\subsection{Pole model\label{subsec:pole_f}}

\begin{figure*}[ht]
\includegraphics[width=\textwidth]{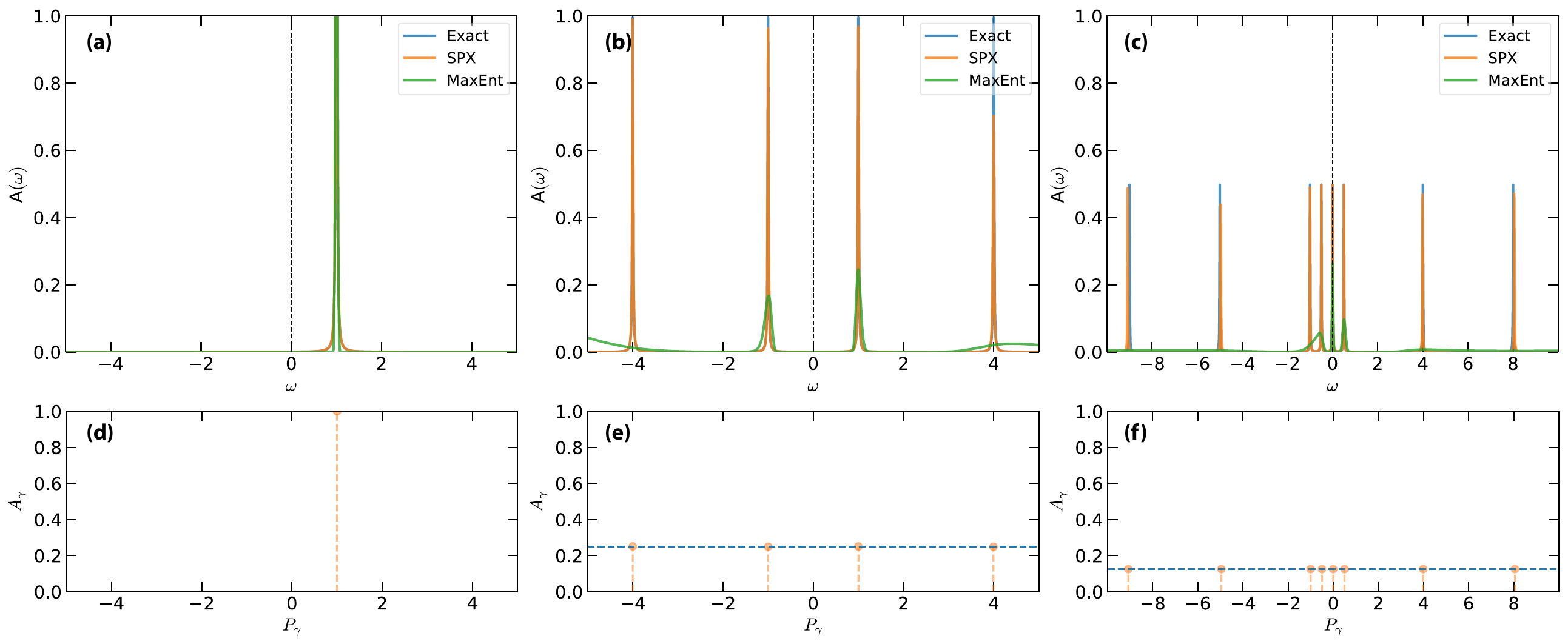}
\caption{Analytic continuations of fermionic correlators (M$_{02}$: Pole model). (a)-(c) Exact and calculated spectral functions. The vertical dashed lines denote the Fermi level. (d)-(f) Distributions of poles for the ``best'' solutions. The horizontal dashed lines denote the exact values of amplitudes of the poles. \label{fig:X06}}
\end{figure*}

\begin{figure*}[ht]
\centering
\includegraphics[width=\textwidth]{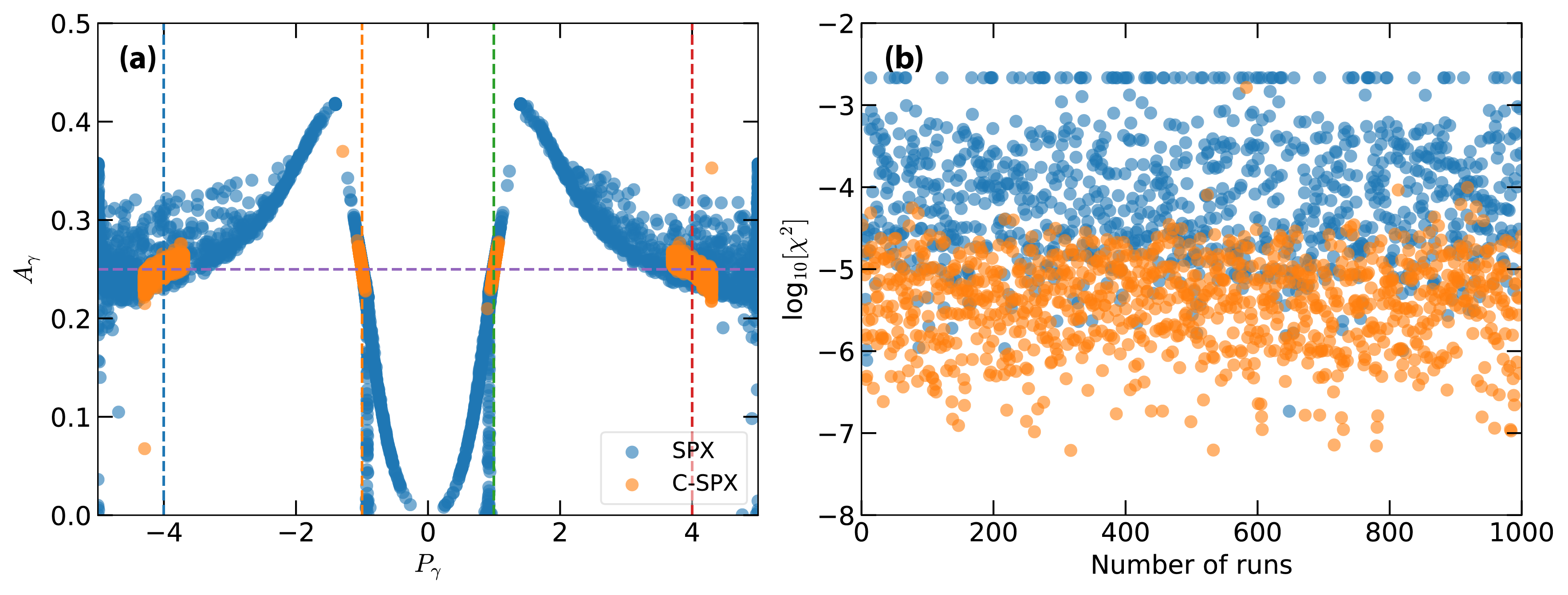}
\caption{Test of the C-SPX method (M$_{02}$: Pole model). A four-pole model is considered. In the C-SPX calculation, the locations of the four poles are restricted in a narrow region, while their amplitudes are free of limitations. The number of individual Monte Carlo runs is 1000. (a) Distribution of the poles. Here the horizontal dashed line means the exact amplitudes of the poles, while the vertical dashed lines denote the exact locations of the poles. (b) Goodness-of-fit function $\chi^{2}$ for repeated Monte Carlo runs. \label{fig:V78}}
\end{figure*}

\begin{figure*}[ht]
\centering
\includegraphics[width=\textwidth]{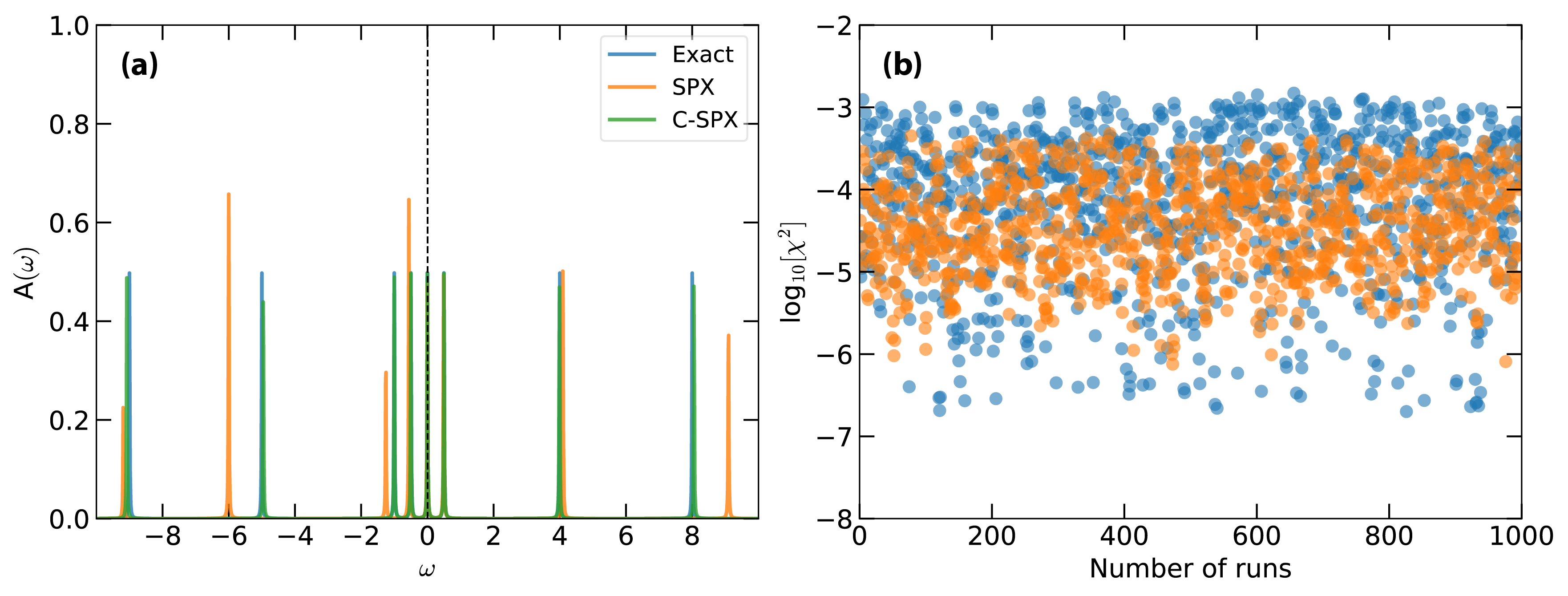}
\caption{Test of the C-SPX method (M$_{02}$: Pole model). An eight-pole model is considered. In the C-SPX calculation, the amplitudes of the eight poles are fixed to 0.125, while the locations are optimized. The number of individual Monte Carlo runs is 1000. (a) Calculated and exact spectral functions. (b) Goodness-of-fit function $\chi^{2}$ for repeated Monte Carlo runs. \label{fig:V09}}
\end{figure*}

Now let us turn to the molecule cases, in which the spectral functions usually exhibit multiple discrete $\delta$-like peaks~\cite{PhysRevB.104.165111,YING2022111549}. To construct the Matsubara data, Eq.~(\ref{eq:pole_f}) is utilized ($\beta = 20.0$). Random noises are added to the synthetic Matsubara data as described above.    

Three typical scenarios are prepared to examine the SPX method: (i) an off-center $\delta$-like peak ($N_p = 1$, $A_1 = 1.0$, $P_1 = 1.0$), (ii) four low-frequency $\delta$-like peaks with equal amplitudes ($N_p = 4$, $A_1 = A_2 = A_3 = A_4 = 0.25$, $P_1 = -P_2 = 4.0$, $P_3 = -P_4 = 1.0$), and (iii) eight $\delta$-like peaks distributed over a wide frequency range ($N_p = 8$, $A_1 = A_2 = A_3 = A_4 = A_5 = A_6 = A_7 = A_8 = 0.125$, $P_1 = 8.0$, $P_2 =4.0$, $P_3 = -P_5 = 0.5$, $P_4 = 0.0$, $P_6 =-1.0$, $P_7 =-5.0$, $P_8 =-9.0$). In the calculations, the number of poles is considered as \emph{a priori} knowledge. Besides Eq.~(\ref{eq:sum_rule_f}), there are no additional limitations for the amplitudes and locations of the poles. Especially, to obtain reasonable solutions for scenarios (ii) and (iii), the numbers of individual Monte Carlo runs are increased to $2 \times 10^5$. Figure~\ref{fig:X06} shows the analytic continuation results. Clearly, both the amplitudes and locations of the peaks (poles) are resolved accurately by the SPX method. Not only the low-frequency multiplets but also the high-frequency sharp peaks are well reproduced. For the three scenarios, the traditional analytic continuation methods, such as MaxEnt~\cite{JARRELL1996133,PhysRevB.44.6011}, fail to distinguish the multiple peaks. They can only give rise to a envelop curve for the $\delta$-like peaks.

Though the three pole models are well solved by the SPX method, quite a lot of computational resources are consumed, especially for scenarios (ii) and (iii). Just as mentioned before, extra constraints could be imposed within the parameterization of the poles, which is a widely used trick for the SAC method and its variants~\cite{PhysRevB.57.10287,beach2004,PhysRevB.62.6317,PhysRevE.94.063308,PhysRevB.76.035115,PhysRevE.81.056701,PhysRevX.7.041072,PhysRevB.101.085111,SHAO20231,PhysRevB.102.035114,PhysRevB.78.174429}. Such constraints, which represent some kind of innate knowledge [e.g., existence of the sharp band edge, and properties of the poles (including number, amplitudes, and approximate locations)], could significantly improve fidelity of the calculated spectrum and computational efficiency of the SPX method. 

The C-SPX method is at first examined by using the four-pole model [i.e. scenario (ii)]. Because the analytic continuation calculation without extra constraints has been done before (see Figure~\ref{fig:X06}), we try to limit the locations of the poles with $P_{\gamma} \in [-4.5,-3.5] \bigcup [-1.5,-0.5] \bigcup [0.5,1.5] \bigcup [3.5,4.5]$ and rerun the calculations (We will discuss the tricks about how to speculate the restricted zones for the poles in Section~\ref{sec:application_b}). The number of individual Monte Carlo is reduced to 1000. The calculated results are displayed in Figure~\ref{fig:V78}. Figure~\ref{fig:V78}(a) shows the distribution of the poles. The horizontal and vertical dashed lines denote the exact amplitudes and locations of the poles, respectively. Figure~\ref{fig:V78}(b) shows the collected $\chi^{2}$ for individual Monte Carlo runs. We find that once the constraints are imposed, the computational accuracy of not only the locations but also the amplitudes of the poles is greatly improved, and it becomes easier to figure out the global minimization. 

And then the eight-pole model [i.e. scenario (iii)] is used to test the C-SPX method. The amplitudes of the eight poles are the same, but they reside in a wide frequency range. It is rather difficult to get the correct solution by using the traditional analytic continuation methods. As for the SPX method, it is tough to arrive at the global minimum unless the number of individual Monte Carlo runs is increased up to $10^6$, which is very time-consuming. However, if we assume that the amplitudes of the poles are known and try to optimize their locations by the C-SPX method, it is easy to reproduce the correct spectrum. Now the number of individual Monte Carlo runs can be reduced to 1000. Figure~\ref{fig:V09} shows the benchmark results. For the SPX method, though the peaks at the low-frequency region are roughly resolved, it fails to reproduce the high-frequency peaks. However, by using the C-SPX method, both the high-frequency and low-frequency peaks are accurately resolved. Furthermore, we find that the values of the goodness-of-fit function $\chi^{2}$ are more scattered when the SPX method is used [See Fig.~\ref{fig:V09}(b)]. It implies that the SPX method is easier trapped by the local minima than the C-SPX method.   

\subsection{Resonance model\label{subsec:resonance_f}}

\begin{figure*}[ht]
\centering
\includegraphics[width=\textwidth]{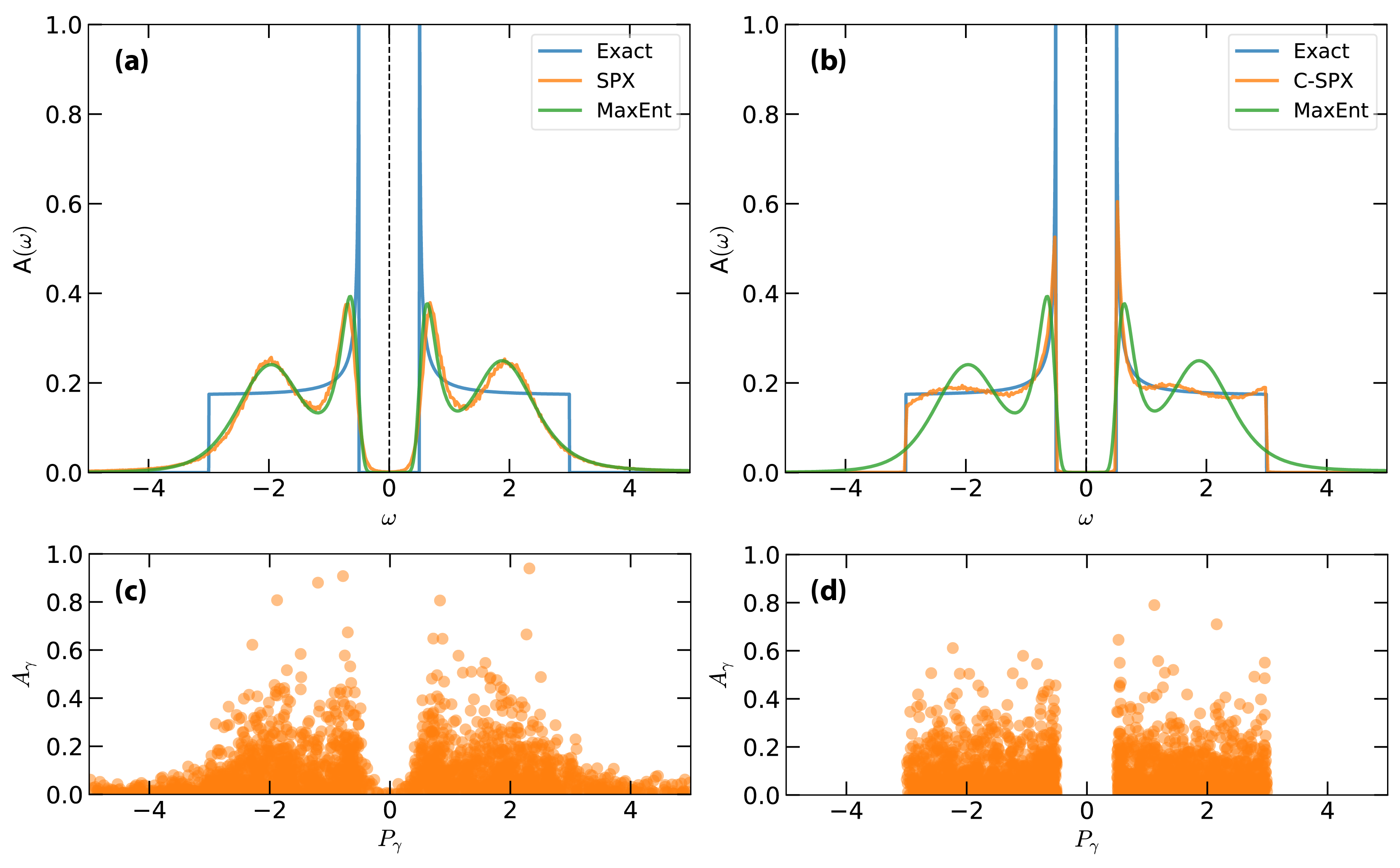}
\caption{Analytic continuations of fermionic correlators (M$_{03}$: Resonance model). The exact and calculated spectral functions are shown in the first row, while the corresponding distributions of poles are shown in the second row. (a) and (c) Results obtained by the SPX method. (b) and (d) Results obtained by the C-SPX method. In the C-SPX calculations, the locations of the poles are limited in $[-3.0,0.5] \bigcup [0.5,3.0]$. \label{fig:X04}}
\end{figure*}

The third example concerns the analytic continuation of Matsubara Green's function of a BCS superconductor~\cite{beach2004}. Its spectral function reads:  
\begin{equation}
\label{eq:model03}
A(\omega) =
\left\{
    \begin{array}{lr}
    \frac{1}{W} \frac{|\omega|}{\sqrt{\omega^2 - \Delta^2}}, ~\quad & \text{if}~\Delta < |\omega| < W/2. \\
    0, & \text{otherwise}.
    \end{array}
\right.
\end{equation}
Here, $W$ denotes the bandwidth ($W = 6.0$), and $\Delta$ is used to control the gap's size ($\Delta = 0.5$). This spectrum is comprised of flat shoulders, steep peaks, and sharp gap edges. These unusual features pose severe challenges to the existing analytic continuation methods. Figure~\ref{fig:X04}(a)-(b) shows the analytic continuation results obtained by the SPX method without any constraints. Clearly, the calculated spectrum exhibits extra shoulder peaks around $\pm 2.0$ and long tails for $|\omega| > 3$. The energy gap and the sharp band edges are not well captured. Then constraints are applied to the locations of the poles. They are allowed to appear in a restricted frequency range ($[-3.0,0.5] \bigcup [0.5,3.0]$). The calculated results are shown in Fig.~\ref{fig:X04}(c)-(d). It is clear that the major characteristics of the exact spectrum are well reproduced by the C-SPX method. The excrescent peaks around $\pm 2.0$ are mostly suppressed, leaving small ripples. 

\subsection{Matsubara Green's function\label{subsec:green_f}}

\begin{figure*}[ht]
\centering
\includegraphics[width=\textwidth]{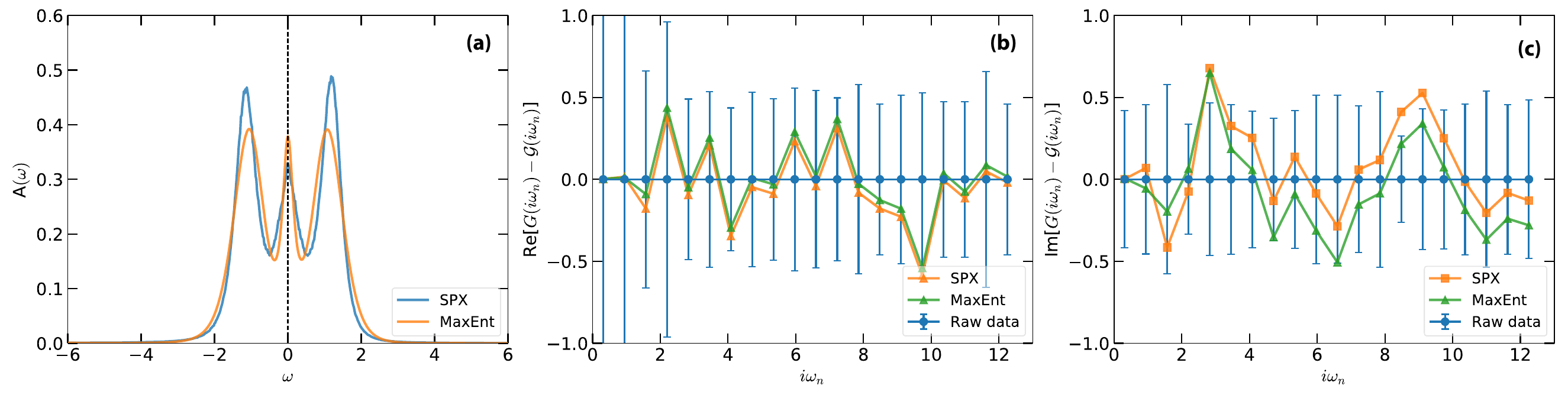}
\caption{Analytic continuations of fermionic correlators (M$_{04}$: Matsubara Green's function). (a) Calculated spectral functions. (b)-(c) Distances between the reproduced Green's function $G(i\omega_n)$ and the raw Green's function $\mathcal{G}(i\omega_n)$. In panels (b) and (c), all the data are scaled by a factor of $10^3$ for a better view. The error bars of $\mathcal{G}(i\omega_n)$ are also shown. \label{fig:X05G}}
\end{figure*}

In this subsection, let us concentrate on a realistic case, Matsubara Green's function from quantum many-body simulation. We just consider the following single-band half-filling Hubbard model:
\begin{equation}
\label{eq:hubbard}
H = -t \sum_{\langle ij \rangle,~\sigma} c^{\dagger}_{i\sigma}c_{j\sigma}
 - \mu \sum_i n_i + U \sum_i n_{i\uparrow} n_{i\downarrow},
\end{equation}
where $t$ is the hopping parameter ($t = 0.5$), $\mu$ is the chemical potential ($\mu = 1.0$), $U$ is the Coulomb interaction ($U = 2.0$), $n$ is the occupation number ($n = 1.0$), $\sigma$ denotes the spin, $i$ and $j$ are site indices. The inverse system temperature is $\beta = 10.0$. This model is defined in a Bethe lattice. It was solved by the single-site dynamical mean-field theory (dubbed DMFT)~\cite{RevModPhys.68.13} in combination with the hybridization expansion continuous-time quantum Monte Carlo impurity solver (dubbed CT-HYB)~\cite{RevModPhys.83.349}. All the calculations were done by using the \texttt{iQIST} software package~\cite{HUANG2015140,HUANG2017423}. The major outputs of the DMFT + CT-HYB calculations are the Green's function and the self-energy function at the imaginary axis. In this example, the Matsubara Green's function is analytically continued to extract the spectral function by the MaxEnt method and the SPX method. The Matsubara self-energy function will be treated in the next subsection.  

The analytic continuation results are shown in Figure~\ref{fig:X05G}. Since the Coulomb interaction is comparable with the bandwidth ($W = 4t = 2.0$), the ground state of this model should be metallic, but close to the Mott metal-insulator transition~\cite{RevModPhys.70.1039}. Thus, it is not strange that the spectral function exhibits a three-peak structure (i.e. the quasiparticle resonance peak in the vicinity of the Fermi level + lower and upper Hubbard bands), which is a hallmark of the strongly correlated metallic systems~\cite{RevModPhys.68.13}. We can observe that the spectral functions given by the MaxEnt method and the SPX method [see Fig.~\ref{fig:X05G}(a)] match with each other, so the results should be reliable. Figure~\ref{fig:X05G}(b)-(c) show the real and imaginary parts of the reproduced Green's functions, which are compared with the input data. Apparently, the original Matsubara data are well reproduced, which implies the derived pole expansion formula [see Eq.~(\ref{eq:pole_f})] is reasonable.

\subsection{Matsubara self-energy function\label{subsec:self_f}}

\begin{figure*}[ht]
\centering
\includegraphics[width=\textwidth]{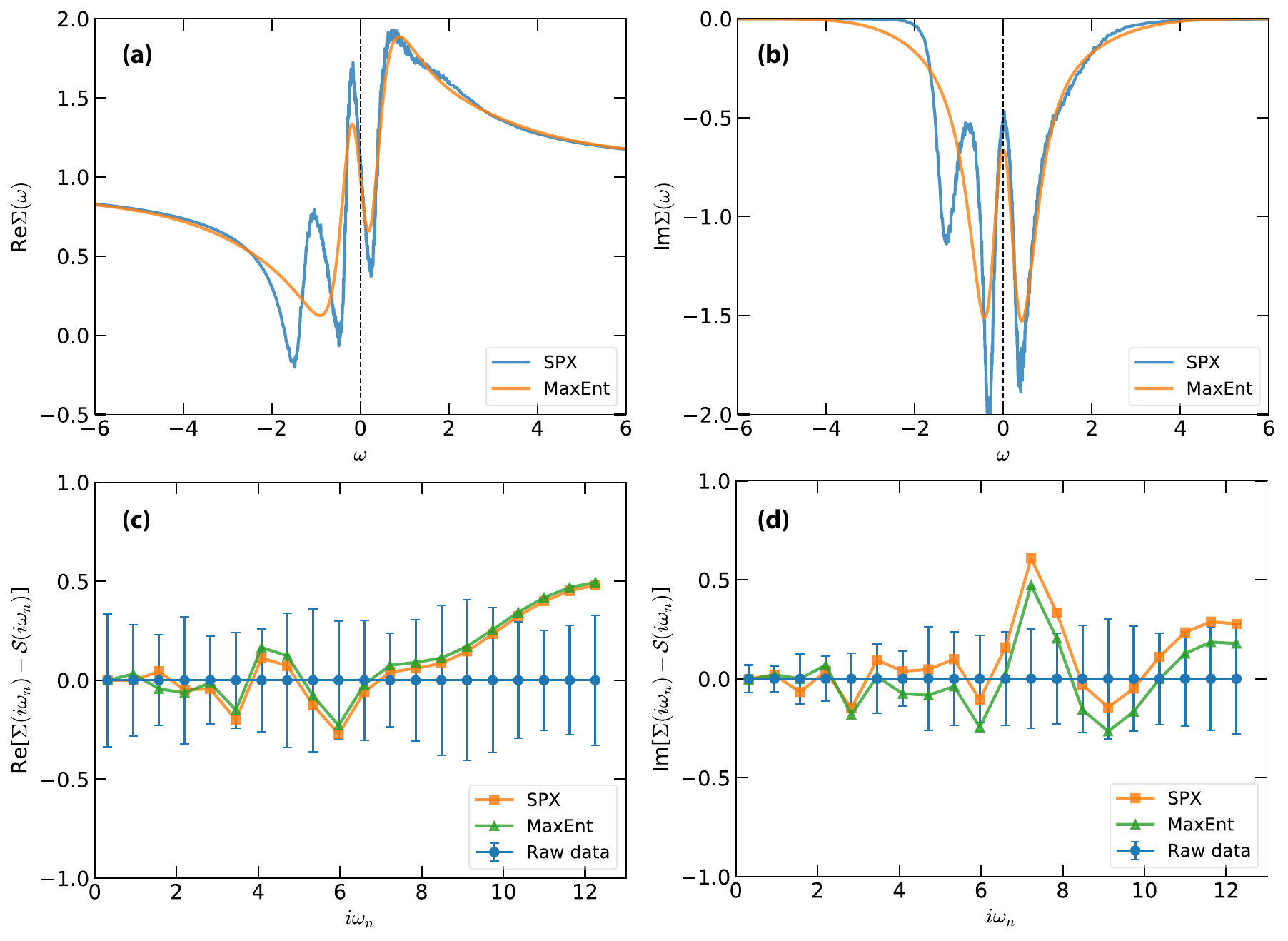}
\caption{Analytic continuations of fermionic correlators (M$_{05}$: Matsubara self-energy function). (a)-(b) Self-energy function at real axis, $\Sigma(\omega)$. (c)-(d) Distances between the reproduced self-energy function $\Sigma(i\omega_n)$ and the raw self-energy function $\mathcal{S}(i\omega_n)$. In panels (c) and (d), all the data are scaled by a factor of $10^2$ for a better view. The error bars of $\mathcal{S}(i\omega_n)$ are also shown. \label{fig:X05S}}
\end{figure*}

Besides one-particle Green's functions, it often is necessary to analytically continue the self-energy functions after the DMFT simulations~\cite{RevModPhys.68.13}. Now we would like to demonstrate how to convert the self-energy function from Matsubara to real frequencies by using the SPX method. The high-frequency asymptotic behaviors of self-energy are different from those of one-particle Green's function. When approaching the high-frequency limit, the real part is a non-zero constant, and the imaginary part decays like $\Sigma_1 /(i\omega_n)$ where $\Sigma_1$ is the first moment of the self-energy~\cite{KAUFMANN2023108519}. So the input data of self-energy should be preprocessed. The original data are taken from the DMFT solution of Eq.~(\ref{eq:hubbard}). At first, the Hartree term $\Sigma_{H}$ is subtracted from $\Sigma(i\omega_n)$:
\begin{equation}
\tilde{\Sigma}(i\omega_n) = \Sigma(i\omega_n) - \Sigma_{H}.
\end{equation}
Note that $\Sigma_{H}$ is approximately equal to the asymptotic value of the real part of $\Sigma(i\omega_n)$ when $n$ goes to infinite. It is also the zeroth moment of the self-energy. Sometimes $\tilde{\Sigma}(i\omega_n)$ could be further normalized by division through its first moment $\Sigma_1$~\cite{KAUFMANN2023108519,PhysRevB.96.155128},
\begin{equation}
\tilde{\Sigma}(i\omega_n) = \frac{\Sigma(i\omega_n) - \Sigma_{H}}{\Sigma_1},
\end{equation}
but it is not necessary. Then, we perform analytic continuation by using the SPX method as usual and take $\tilde{\Sigma}(i\omega_n)$ as the input data. The analytic continuation procedure will convert $\tilde{\Sigma}(i\omega_n)$ into $\tilde{\Sigma}(\omega)$. Finally, $\tilde{\Sigma}(\omega)$ is supplemented with the Hartree term $\Sigma_{H}$ to get $\Sigma(\omega)$:
\begin{equation}
\Sigma(\omega) = \tilde{\Sigma}(\omega) + \Sigma_{H}.
\end{equation}

The benchmark results are illustrated in Fig.~\ref{fig:X05S}. The results obtained by the MaxEnt method~\cite{JARRELL1996133,PhysRevB.44.6011} are also displayed as a comparison. When $\omega > 0$, the results obtained by both methods are consistent with each other. If $\omega < 0$, the results obtained by the SPX method exhibit an additional peak near $\omega = -2.0$ [see Fig.~\ref{fig:X05S}(a)-(b)]. On the contrary, the MaxEnt method only yields a smooth spectrum. Anyway, the reproduced Matsubara data generated by both methods are quite close to the original Matsubara data [see Fig.~\ref{fig:X05S}(c)-(d)].    

\section{Applications: Bosonic correlators\label{sec:application_b}}

In this section, we are going to test whether the SPX method supports analytic continuations of bosonic correlators. Traditionally, the analytic continuations of single-particle Green's functions have attracted the most attention~\cite{JARRELL1996133}. This is because the obtained single-particle spectral functions can be easily observed by photoemission spectroscopy. Over the last ten years, advances in quantum many-body theories have made it possible to study strongly correlated electron systems beyond the single-particle level~\cite{PhysRevB.105.245104,PhysRevX.11.041040,PhysRevB.101.165105,PhysRevB.101.075122}. The two-particle quantities have become more and more important, since they are the key building blocks in these newly established many-body computational methods~\cite{PhysRevB.99.115112,PhysRevB.90.235135,PhysRevB.92.081106}. Similar to the single-particle Green's functions, the two-particle quantities can not be compared directly with the experiments. We have to find a reliable method for the analytic continuations of two-particle functions and get the corresponding dynamical susceptibilities, which can be indeed measured by experiments. Since the two-particle quantities are usually formulated on the bosonic Matsubara frequencies~\cite{PhysRevB.84.075145,PhysRevB.97.205111,PhysRevB.96.035147}, the analytic continuation methods for fermionic correlators should be modified to meet this requirement. This is not always a trivial task~\cite{PhysRevB.94.245140}. In this section, we demonstrate that the SPX method can, with the appropriate kernels, symmetry relations, and constraints, also be used for bosonic functions. Six examples are presented here. The benchmark results indicate that the SPX method is able to treat not only synthetic but also realistic two-particle correlation functions.      

\subsection{Gaussian model\label{subsec:gauss_b}}

\begin{figure}[t]
\includegraphics[width=\columnwidth]{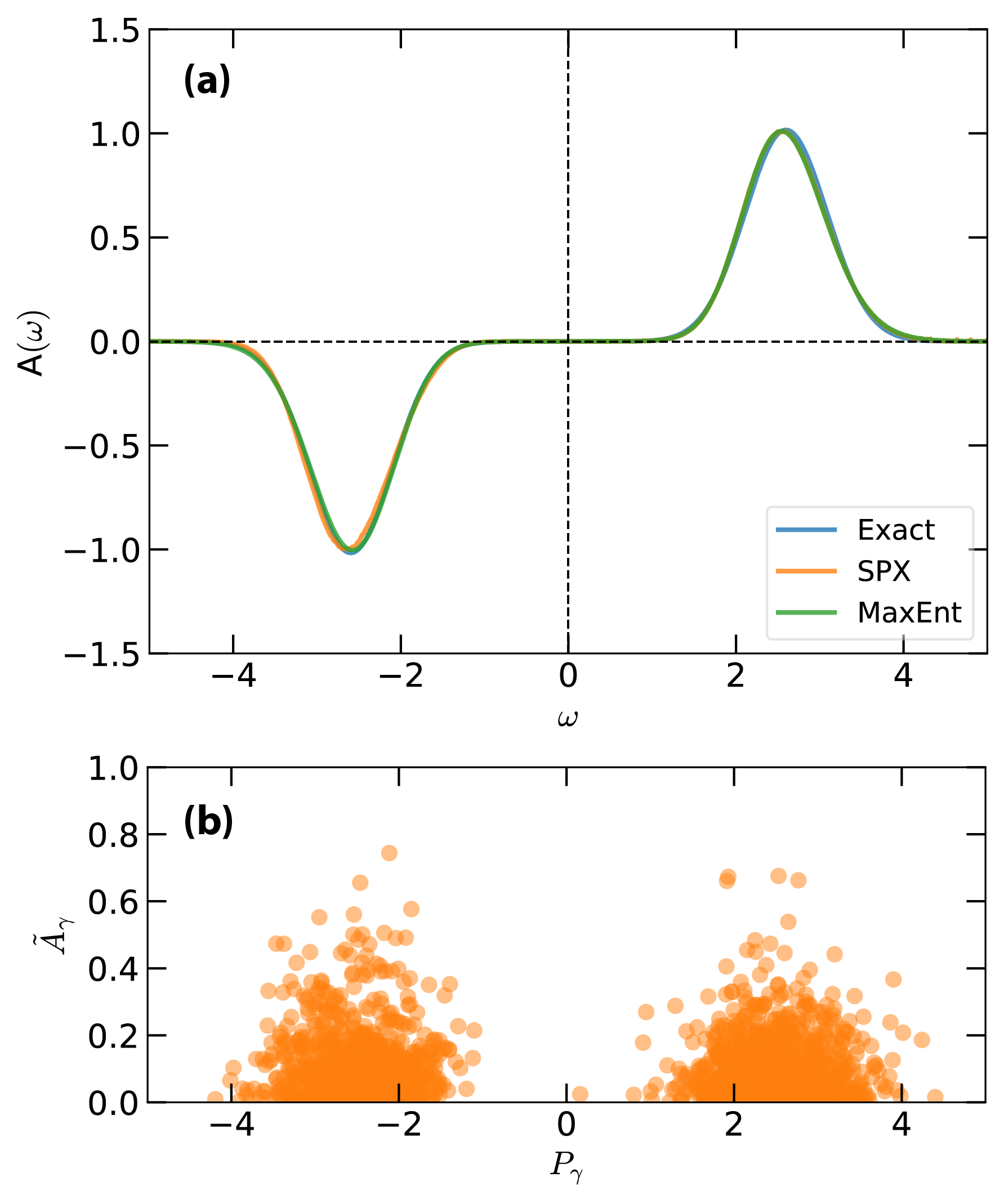}
\caption{Analytic continuations of bosonic correlators (M$_{06}$: Gaussian model). (a) Exact and calculated spectral functions. The vertical dashed line denotes the Fermi level. (b) Distribution of poles for a ``good'' solution as gathered in the SPX simulation. Notice that the weights of these poles are rescaled for a better view. \label{fig:Y01}}
\end{figure}

The first example we address here is a Gaussian model. The analytic structure of the exact spectral function is:
\begin{equation}
\label{eq:model05}
\frac{A(\omega)}{\omega} =
 \alpha_1 \exp\left[-\frac{(\omega - \epsilon_1)^2}{2\gamma^2_1}\right] +
 \alpha_2 \exp\left[-\frac{(\omega - \epsilon_2)^2}{2\gamma^2_2}\right],
\end{equation}
where $\alpha_1 = \alpha_2 = 0.5$, $\gamma_1 = \gamma_2 = 0.5$, $\epsilon_1 = -\epsilon_2 = 2.5$. Clearly, $A(\omega)$ is an odd function. It exhibits two distinct peaks at energies $\epsilon_1$ and $\epsilon_2$. And the two Gaussian peaks are antisymmetric about $\omega = 0.0$. There is a huge gap between the two peaks ($\Delta_{\text{gap}} \approx 3.0$). The Matsubara data are generated by using Eqs.~(\ref{eq:spectral_b}) and (\ref{eq:kernel_b}). The inverse system temperature $\beta$ is 10.0.   

In Fig.~\ref{fig:Y01}(a), the exact spectrum is drawn and compared to the calculated spectra as obtained by the SPX and MaxEnt methods. Both methods could capture the major characteristics, including the two-peaks structure and the big gap, of the spectrum. The lower panel of Fig.~\ref{fig:Y01} shows a typical distribution of the poles. The overall profile of this distribution is also Gaussian-like. There are a small fraction of poles in the band gap region, but their contributions are trivial.

\subsection{Pole model\label{subsec:pole_b}}

\begin{figure}[t]
\includegraphics[width=\columnwidth]{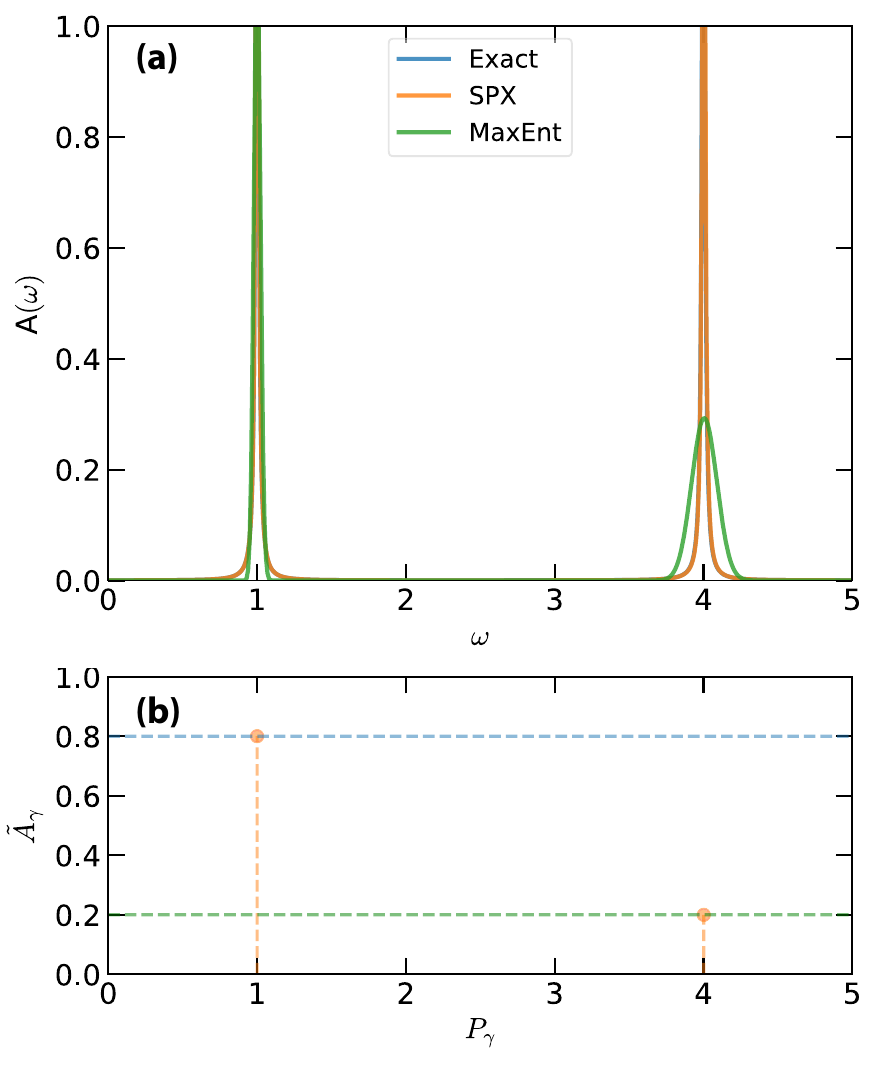}
\caption{Analytic continuations of bosonic correlators (M$_{07}$: Pole model). Only the spectra in the positive half-axis are shown. (a) Exact and calculated spectral functions. (b) Distribution of poles for the ``best'' solution as collected in the SPX simulation. The horizontal dashed lines mark the exact amplitudes of the poles. \label{fig:Y05}}
\end{figure}

For the second example, we consider a four-pole model. Its analytic form is as follows:
\begin{equation}
\label{eq:model06}
G(z) = \sum^{4}_{i=1} \frac{\alpha_i}{z - \epsilon_i},
\end{equation}
where $\alpha_1 = \alpha_3 = 0.5$, $\alpha_2 = \alpha_4 = -0.5$, $\epsilon_1 = -\epsilon_2 =  4.0$, and $\epsilon_3 = -\epsilon_4 = 1.0$. The exact spectrum exhibits four off-centered $\delta$-like peaks at $\epsilon_1$, $\epsilon_2$, $\epsilon_3$, and $\epsilon_4$. These peaks are also antisymmetric with respect to $\omega = 0.0$. Note that similar spectra are usually seen in molecule systems (such as the Hubbard dimer) and the momentum-resolved Kohn-Sham eigenvalues (i.e. the ``band structure'') of condensed matter systems~\cite{PhysRevLett.126.056402,PhysRevB.107.075151}. We just use this equation or Eq.~(\ref{eq:pole_b}) to generate the Matsubara data. The inverse system temperature $\beta$ is 20.0.

In this example, we respect the symmetries of the Green's function and spectral function. In other words, $G(z)$ is treated as a bosonic correlator of a Hermitian operator, and the corresponding spectrum can be defined in the positive half-axis only. The exact and calculated spectral functions are shown in Fig.~\ref{fig:Y05}(a). As for the MaxEnt method, it resolves the low-frequency peak at $\omega = 1.0$ quite well. However, it fails to resolve the high-frequency peak at $\omega = 4.0$. The resulting high-frequency peak is much broader and smoother than the exact one. On the other hand, the SPX method does an excellent job. Not only the positions but also the heights of the two $\delta$-like peaks are precisely reproduced. In Fig.~\ref{fig:Y05}(b) the poles belonged to the ``best'' solution are visualized. Clearly, the weights and locations of the poles agree quite well with the exact values.      

\subsection{Resonance model\label{subsec:resonance_b}}

\begin{figure}[t]
\includegraphics[width=\columnwidth]{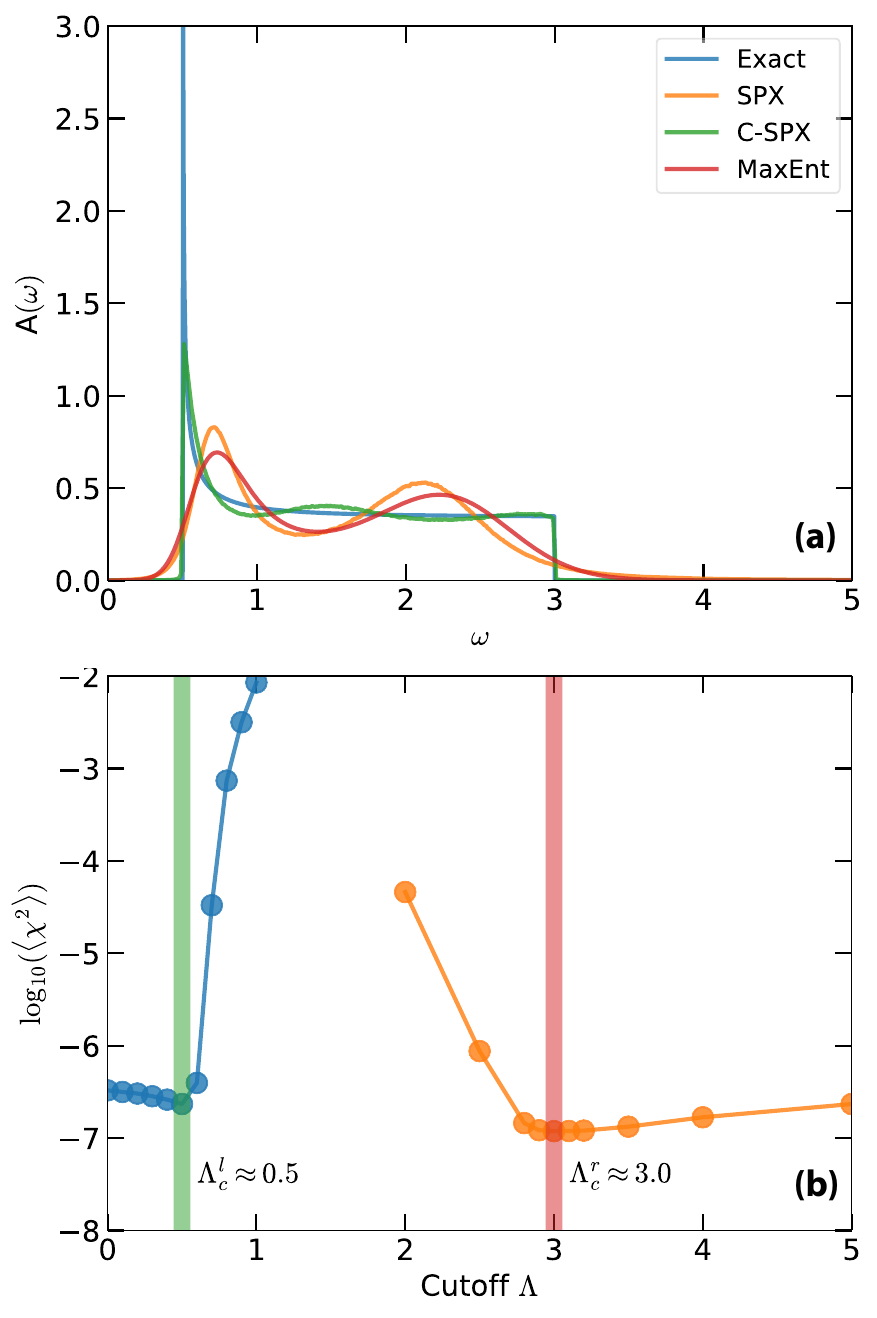}
\caption{Analytic continuations of bosonic correlators (M$_{08}$: Resonance model). (a) Exact and calculated spectral functions. (b) Goodness-of-fit function $\chi^2$ with respect to two artificial cutoff parameters $\Lambda^{l}$ and $\Lambda^{r}$, which are used to restrict the possible locations of the poles. Here, the two vertical bars denote the optimal cutoffs $\Lambda^{l}_c$ and $\Lambda^{r}_c$, which are approximately 0.5 and 3.0, respectively. \label{fig:Y09}}
\end{figure}

For the third example, we consider a resonance model. Its spectral function is defined as follows: 
\begin{equation}
\label{eq:model07}
A(\omega) =
\left\{
    \begin{array}{lr}
    \frac{1}{W} \frac{\omega}{\sqrt{\omega^2 - \Delta^2}}, ~\quad & \text{if}~\Delta < \omega < W/2. \\
    0, & \text{otherwise}.
    \end{array}
\right.
\end{equation}
It is obviously a variant of Eq.~(\ref{eq:model03}). The $W$ and $\Delta$ parameters are the same with those used in M$_{03}$ (see Section~\ref{subsec:resonance_f}). This spectrum has finite weights only at the positive half-axis. It should exhibit a sharp resonance peak at $\omega = \Delta$ and a broad platform [see Fig.~\ref{fig:Y09}(a)]. We try to generate the bosonic Green's function via Eqs.~(\ref{eq:spectral_h}) and (\ref{eq:kernel_h}). The inverse system temperature is $\beta = 10.0$.

As usual, both the MaxEnt and SPX methods are employed to solve the analytic continuation problem. The results are visualized in Fig.~\ref{fig:Y09}(a). We observe that the simulated spectra exhibit two broad humps at $\omega \approx$ 0.8 and 2.1, respectively. However, none of the important features of the ideal spectrum is captured. Therefore we resort to the C-SPX method again. Since there are two band edges in the true spectrum, we introduce two cutoff parameters, namely $\Lambda^{l}$ and $\Lambda^{r}$, where $\omega_{\text{min}} < \Lambda^{l} < \Lambda^{r} < \omega_{\text{max}}$. Here, the superscripts ``$l$'' and ``$r$'' mean the left and right band edges, respectively. Thus, the forbidden region for the poles becomes $[\omega_{\text{min}},\Lambda^{l}] \bigcup [\Lambda^{r},\omega_{\text{max}}]$. Next, we try to calibrate the two parameters and evaluate the corresponding $\chi^{2}(\Lambda^{l})$ and $\chi^{2}(\Lambda^{r})$. The calculated results are presented in Fig.~\ref{fig:Y09}(b). We find that neither $\chi^{2}(\Lambda^{l})$ nor $\chi^{2}(\Lambda^{r})$ is monotonic function. They both exhibit minima: 
\begin{eqnarray}
\Lambda^{l}_{c} &=& \mathop{\arg\min}\limits_{\Lambda^l} \chi^2(\Lambda^l), \\
\Lambda^{r}_{c} &=& \mathop{\arg\min}\limits_{\Lambda^r} \chi^2(\Lambda^r).
\end{eqnarray}
We can conclude that $\Lambda^{l}_c = 0.5$ and $\Lambda^{r}_c = 3.0$. They are the optimal cutoff parameters. Such that the restrictions are fixed. We conduct the analytic continuation simulation with the SPX method again. The simulated spectrum is shown in Fig.~\ref{fig:Y09}(a) as well. It is apparent that the C-SPX method outperforms the SPX method and the MaxEnt method in this example. Both the sharp resonance peak at $\omega = 0.5$ and the broad platform at $1.0 < \omega < 3.0$ are well reproduced. The only deviation is the small ripples in the platform region as seen in the simulated spectrum. But these ripples can be further suppressed by using more poles and collecting more ``good'' solutions.

\subsection{Spin-spin correlation function\label{subsec:chi_b}}

\begin{figure}[t]
\includegraphics[width=\columnwidth]{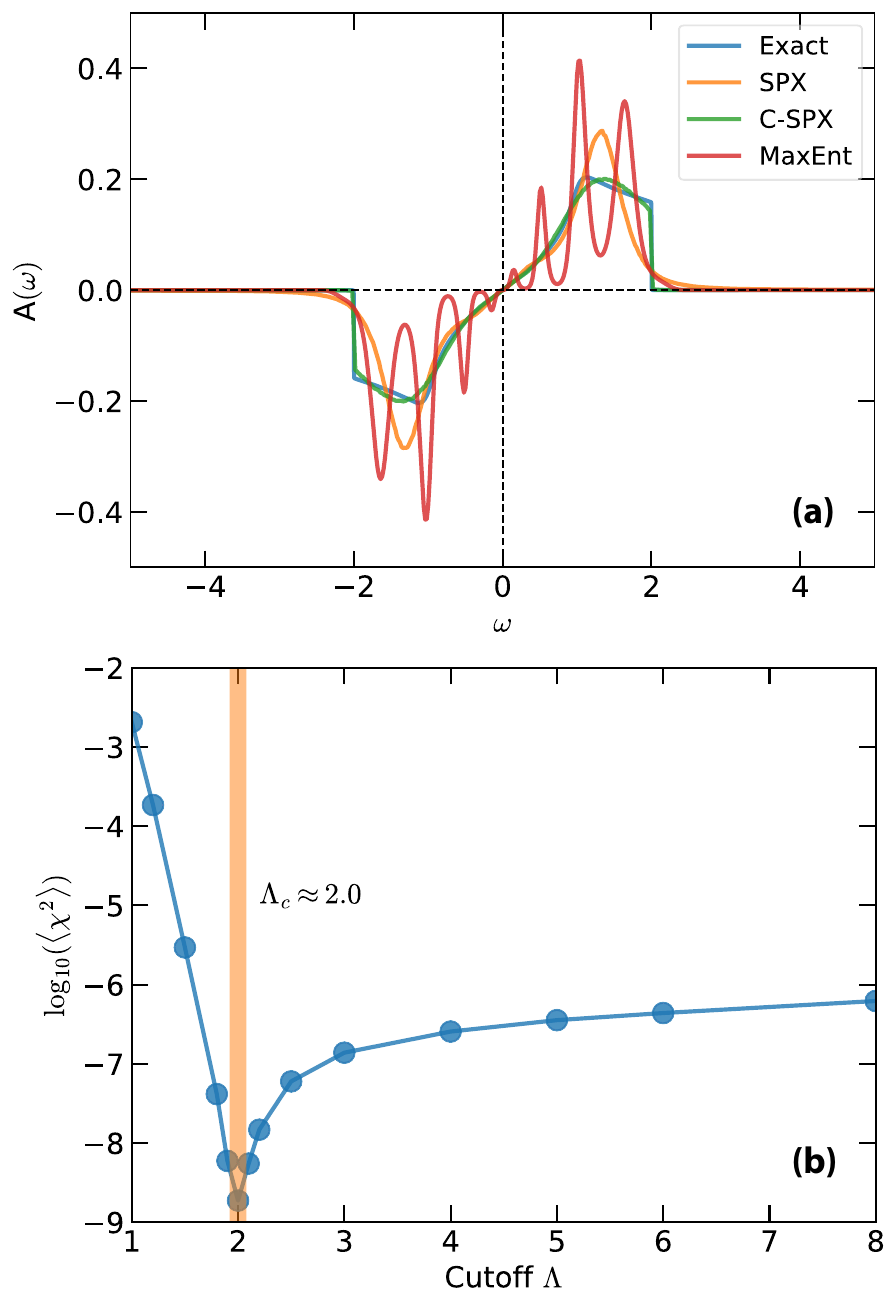}
\caption{Analytic continuations of bosonic correlators (M$_{09}$: Spin-spin correlation function). (a) Exact and calculated spectral functions. (b) Goodness-of-fit function $\chi^2$ with respect to the cutoff parameter $\Lambda$, which is used to limit the left and right thresholds of the spectral function. Here, the vertical bar denotes the optimal cutoff $\Lambda_c$, which is approximately 2.0. \label{fig:SZ20}}
\end{figure}
 
For the fourth example, we consider a spin model in the XY chain. Its Hamiltonian reads:
\begin{equation}
\label{eq:xychain}
H = J_{xy} \sum_{\langle ij \rangle} \left(S^i_x S^j_x  + S^i_y S^j_y \right),
\end{equation}
where $S_\alpha = \frac{\sigma_\alpha}{2}$ are the spin-$\frac{1}{2}$ operators ($\alpha = x,~y,~z$), $\langle ij\rangle$ denotes the nearest neighbors, and $J_{xy} = 1$. This model can be exactly solved by the Jordan-Wigner transformation. The energy spectrum is given by:
\begin{equation}
\epsilon_k = J_{xy}\cos(ka),
\end{equation}
where $a$ is the lattice spacing. The local spin-spin correlation function $\chi_{zz}(\tau) = \langle{S_z(\tau) S_z(0)}\rangle$ is basically a density-density correlator in the sense of spinless fermions. Its exact expression reads~\cite{katsura1970dynamical}:
\begin{equation}
\chi_{zz}(\tau) =\int_{-\pi}^{\pi}
    \frac{\mathrm{d} k \mathrm{d} k'}{(2\pi)^2} 
    \frac{e^{\tau\epsilon_k} e^{(\beta-\tau)\epsilon_{k'}}}{(1+e^{\beta\epsilon_k})(1+e^{\beta\epsilon_{k'}})}.
\end{equation}
Therefore the correspondingly spectral function is:
\begin{equation}
\label{eq:exact_chi_spectrum}
A(\omega) =\int_{-\pi}^{\pi}
    \frac{\mathrm{d} k \mathrm{d} k'}{2\pi}
    \frac{(1-e^{-\beta \omega}) e^{\beta \epsilon_{k'}}}{(1+e^{\beta\epsilon_k})(1+e^{\beta\epsilon_{k'}})} 
    \delta(\omega  - \epsilon_{k'} + \epsilon_k).
\end{equation}
It is obvious that there are thresholds at $\omega = \pm 2J_{xy}$ because  $\max|\epsilon_k - \epsilon_{k'}|$ is $2|J_{xy}|$.

The input Matsubara data $\chi_{zz}(i\omega_n)$ used in the SPX method are generated by the continuous matrix product operator (dubbed cMPO) method~\cite{PhysRevLett.125.170604}. Within this approach, the partition function $Z$ at finite temperature is formulated as a spacetime tensor network living on an infinite cylinder with circumference $\beta = 1/T$. This tensor network is contracted by a boundary continuous matrix product state (dubbed cMPS) with bond dimension $d$ through a process of minimizing free energy. With these in hand, one can get direct access to the local two-time correlator $\chi_{zz}(\tau)$ as well as $\chi_{zz}(i\omega_n)$ in the thermodynamic limit without error bar and time discretization error:
\begin{equation}
\chi_{zz}(\tau)  
= \frac{\mathrm{Tr} \left[ e^{-(\beta - \tau) K_{\wang}} S_z e^{-\tau K_{\wang}} S_z \right]}
       {\mathrm{Tr} e^{- \beta K_{\wang}}},
\end{equation}
and
\begin{equation}
\chi_{zz}(i\omega_n) = \frac{1}{Z}
    \sum_{nm} \left|\langle n| S_z |m \rangle\right|^2
    \frac{e^{-\beta E_n} - e^{-\beta E_m} }{i \omega_n - E_m + E_n}.
\end{equation}
where $K_{\wang}$ is a matrix obtained by contracting the Hamiltonian cMPO and local boundary cMPS, and $E_n$ and $|n \rangle$ are the $n$-th eigenvalue and eigenvector of $K_{\wang}$, respectively. In the present case, we set $\beta = 20.0$ and $d = 16$.

Figure~\ref{fig:SZ20} shows analytic continuation results of the spin-spin correlation function. The exact spectrum is generated by using Eq.~(\ref{eq:exact_chi_spectrum}). As is seen in the upper panel of Fig.~\ref{fig:SZ20}, the MaxEnt method yields a wrong spectrum with strong oscillations. Though these oscillations will be gradually suppressed with increasing temperature, the MaxEnt method has trouble in reproducing the sharp band edges at $|\omega| \approx 2.0$ and the gentle slopes at $1.0 < |\omega| < 2.0$ and $|\omega| < 1.0$. By using the SPX method, the calculated spectrum resembles the exact one as a whole. Especially, the oscillations are absent and the quasi-linear behavior near the Fermi level is well reproduced. However, the calculated spectrum manifests two Gaussian-like peaks around $\omega \approx \pm 1.6$, instead of gentle slopes. Furthermore, there are nontrivial weights in the high-frequency region ($|\omega| > 2.0$).

In order to remedy the above deviations, we have recourse to the C-SPX method again. In this method, the locations of the poles could be restricted. To be more specific, we introduce a new user-supplied cutoff parameter $\Lambda$, which is between $\omega_{\text{min}}$ and $\omega_{\text{max}}$. The forbidden region for the poles is set to $[\omega_{\text{min}},\Lambda] \bigcup [\Lambda,\omega_{\text{max}}]$. Thus, the remaining problem is about how to figure out the optimal $\Lambda$. We just carry out a series of standard SPX calculations with different $\Lambda$ and record the corresponding $\chi^{2}$. Note that the $\chi^{2}(\Lambda)$ function is not monotonic. The optimal $\Lambda$ should be $\Lambda_c = \mathop{\arg \min}_\Lambda \chi^2(\Lambda)$. In the lower panel of Fig.~\ref{fig:SZ20}, $\chi^{2}(\Lambda)$ is shown. This plot can be split into two parts: (i) $\Lambda < 2.0$. $\chi^2$ drops quickly as $\Lambda$ increases. (ii) $\Lambda > 2.0$. $\chi^{2}$ at first increases quickly, and then it approaches its asymptotic value step by step. It is apparent that the $\Lambda_c$ is about 2.0, as indicated by a vertical bar. Now the constraint is fixed. The spectrum obtained by the C-SPX method is shown in Fig.~\ref{fig:SZ20}(a) for a comparison. It agrees well with the exact spectrum. Not only the two thresholds but also the gentle slopes are reproduced perfectly.

\subsection{Current-current correlation function\label{subsec:sigma_b}}

\begin{figure}[t]
\includegraphics[width=\columnwidth]{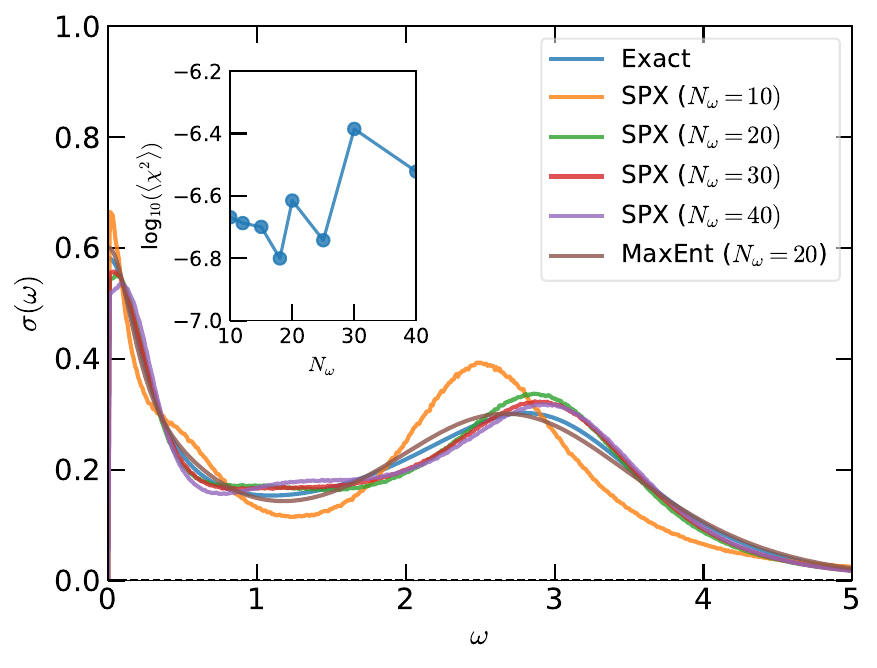}
\caption{Analytic continuations of bosonic correlators (M$_{10}$: Current-current correlation function). Here, the impact of the size of input data ($N_{\omega}$) is studied. The $\chi^2$  as a function of $N_\omega$ is shown in the inset. \label{fig:Y07}}
\end{figure}

In the fifth example, we would like to show how to extract optical conductivity $\sigma(\omega)$ from current-current correlation function $\Pi(i\omega_n)$ by using the SPX method. In imaginary time axis, the current-current correlation function $\Pi(\tau)$ reads:
\begin{equation}
\Pi(\tau) = \frac{1}{3N} \langle \mathbf{j}(\tau) \cdot \mathbf{j}(0) \rangle,
\end{equation}
where $N$ is the number of sites, $\mathbf{j}$ is the current operator, and $\langle ... \rangle$ means the thermodynamic average~\cite{JARRELL1996133}. $\Pi(\tau)$ is a bosonic function. Its spectrum, the frequency dependent optical conductivity $\sigma(\omega)$, is an even function, i.e, $\sigma(\omega) = \sigma(-\omega)$. $\Pi(\tau)$ is related to $\sigma(\omega)$ via the following equation:
\begin{equation}
\Pi(\tau) = \int^{+\infty}_{0} d\omega~K(\tau,\omega) \sigma(\omega),
\end{equation}
Here, the kernel $K(\tau,\omega)$ is already defined by Eq.~(\ref{eq:kernel_h_t}). Since the SPX method needs Matsubara data as input. We should convert $\Pi(\tau)$ to $\Pi(i\omega_n)$ via Fourier transformation. The relation between $\Pi(i\omega_n)$ and $\sigma(\omega)$ reads:
\begin{equation}
\label{eq:current}
\Pi(i\omega_n) = \int^{+\infty}_{0} d\omega~
    K(\omega_n,\omega) \sigma(\omega). 
\end{equation}
The kernel $K(\omega_n,\omega)$ is evaluated by using Eq.~(\ref{eq:kernel_h}). In this example, the analytic expression of $\sigma(\omega)$ is assumed to be:
\begin{equation}
\label{eq:optic}
\sigma(\omega) = \frac{T_1(\omega) + T_2(\omega) + T_3(\omega)}{1 + (\omega/\gamma_3)^6},
\end{equation}
and
\begin{eqnarray}
T_1(\omega) &=& \frac{\alpha_1}{1 + (\omega/\gamma_1)^2},\nonumber\\
T_2(\omega) &=& \frac{\alpha_2}{1 + [(\omega - \epsilon)/\gamma_2]^2},\nonumber\\
T_3(\omega) &=& \frac{\alpha_2}{1 + [(\omega + \epsilon)/\gamma_2]^2},
\end{eqnarray}
where $\alpha_1 = 0.3$, $\alpha_2 = 0.2$, $\gamma_1 = 0.3$, $\gamma_2 = 1.2$, $\gamma_3 = 4.0$, and $\epsilon = 3.0$. This model is borrowed from Ref.~[\onlinecite{PhysRevB.82.165125}]. It manifests two peaks in the positive half-axis. The narrow one at $\omega = 0.0$ is called the Drude peak, which signals a metallic state. Another broad hump is at approximately $\omega = \epsilon$, which is usually from the contributions of interband transitions~\cite{RevModPhys.83.471}. Then the Matsubara data of $\Pi(i\omega)$ is generated by Eq.~(\ref{eq:current}) and Eq.~(\ref{eq:kernel_h}). The inverse system temperature $\beta$ is 20.0. 

Figure~\ref{fig:Y07} shows analytic continuation results by using the SPX method and the MaxEnt method. At first, the Drude peak is very well described. Second, another peak ascribed to the effect of interband transitions, is formed but at a slightly small energy. This is not surprising. In the current SPX simulations, only 10 data points are read as input. The maximum Matsubara frequency is around 2.80, which is smaller than the energy scale of the peak on the real axis ($\omega \approx 3.0$). It is expected that the peak should be described poorly. If more data points are included and more information is added, the theoretical spectrum should be improved. Similar trends have been observed in the previous studies~\cite{PhysRevB.82.165125,PhysRevB.94.245140}. Third, the spectrum by the SPX method exhibits a shoulder peak in the vicinity of $\omega = 0.5$. The origin of this additional peak remains unknown. To examine the size effect of the input dataset, we enlarge the number of input data points up to 40 and perform the calculations again. We find that the interband transition peak is improved, and the shoulder peak is suppressed, but the gain and loss in the goodness-of-fit function are ambiguous.    

\subsection{Lindhard function\label{subsec:lindhard_b}}

\begin{figure*}[ht]
\includegraphics[width=\textwidth]{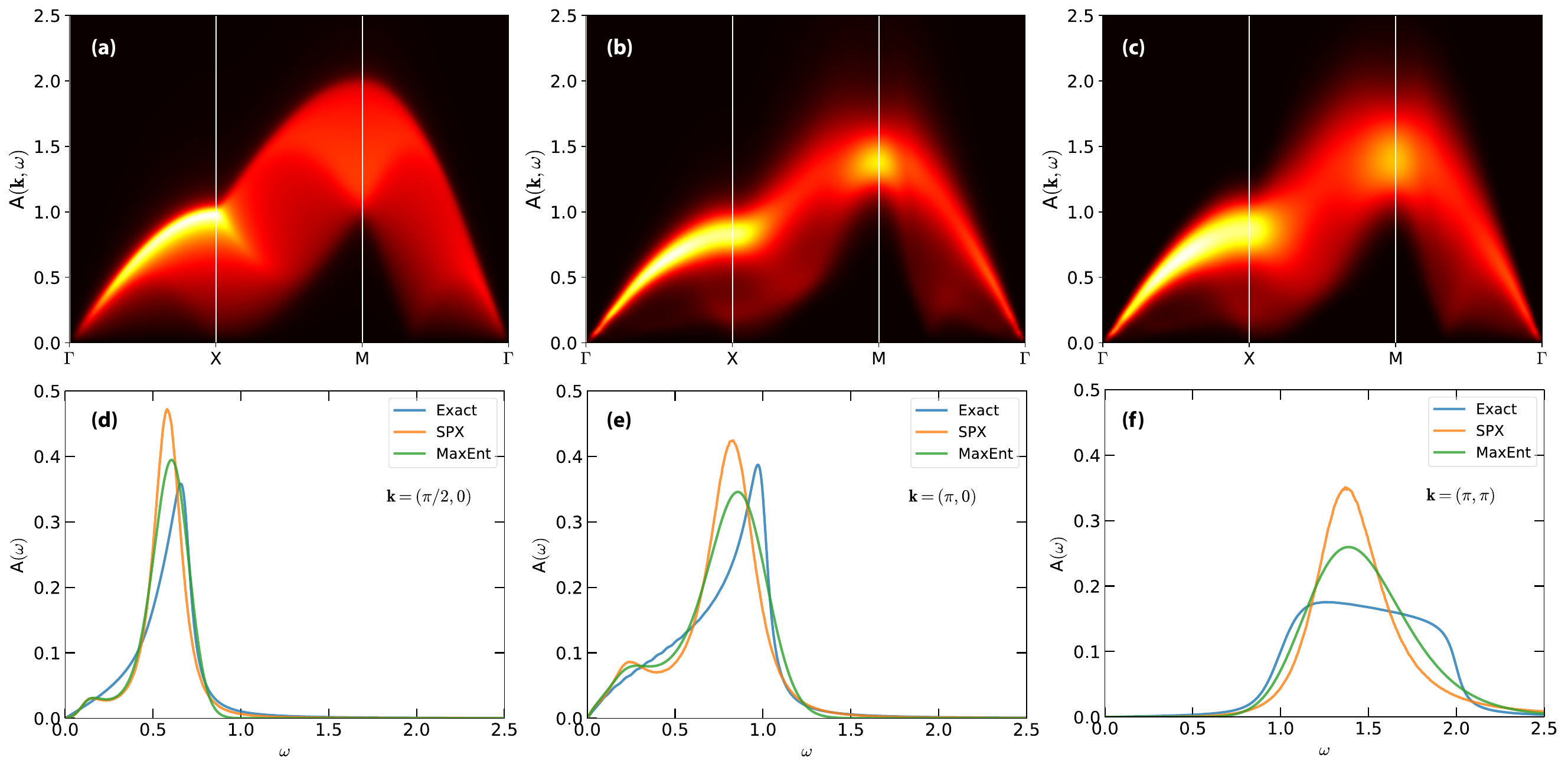}
\caption{Analytic continuations of Lindhard functions of a 2$d$ doped Hubbard model (M$_{11}$). (a)-(c) Momentum-resolved spectral functions A$(\mathbf{k},\omega)$. Panel (a) shows the exact spectrum, while the spectra shown in (b) and (c) are obtained by the SPX method and the MaxEnt method, respectively. (d)-(f) Spectra at some selected high-symmetry points $\mathbf{k}$ in the Brillouin zone. Here $\mathbf{k} = (\pi/2,0)$, $X~(\pi,0)$, and $M~(\pi,\pi)$. \label{fig:Y08}}
\end{figure*}

The Lindhard function represents the basic building block of many-body physics. It describes the charge response of electrons to an external perturbation~\cite{many_body_book,many_body_book_2016}. In this example, we will try to address the analytic continuation problem of the Lindhard function. The analytic expression of the Lindhard function is as follows:
\begin{equation}
\label{eq:lind}
\chi(i\omega_n,\mathbf{q}) = \sum_{\mathbf{p}}
\frac{n_F(\epsilon_{\mathbf{p}}) - n_F(\epsilon_{\mathbf{p} + \mathbf{q}})}
     {i\omega_n + \epsilon_{\mathbf{p}} - \epsilon_{\mathbf{p} + \mathbf{q}}},
\end{equation}
where $\mathbf{p}$ and $\mathbf{q}$ are the wave vectors, $n_{F}(\epsilon)$ means the Fermi-Dirac distribution,
\begin{equation}
n_{F}(\epsilon) = \frac{1.0}{e^{\beta(\epsilon - \mu)} + 1.0},
\end{equation}
 and $\epsilon_{\mathbf{p}}$ denotes the band dispersion. Since we consider a doped tight-binding model on a square lattice with the nearest neighbor hopping,
\begin{equation}
\epsilon_{\mathbf{p}} = -2t [\cos{(p_x)} + \cos{(p_y)}].
\end{equation}
Here, $t$ is the hopping parameter ($t = 0.25$). The chemical potential $\mu$ is set to -0.5, and the inverse temperature is set to 50.0. Note that this model is the same with the one as used in Ref.~[\onlinecite{PhysRevB.94.245140}]. The momentum-resolved spectral functions should be evaluated by -Im $\chi^{R}(\omega, \mathbf{q}) / \pi$, where  
\begin{equation}
\label{eq:hard}
\chi^{R}(\omega,\mathbf{q}) = \sum_{\mathbf{p}}
\frac{n_F(\epsilon_{\mathbf{p}}) - n_F(\epsilon_{\mathbf{p} + \mathbf{q}})}
     {\omega + i\eta + \epsilon_{\mathbf{p}} - \epsilon_{\mathbf{p} + \mathbf{q}}}.
\end{equation}
We at first try to calculate $\chi(i\omega_n,\mathbf{q})$ along the selected high-symmetry directions ($\Gamma - X - M - \Gamma$) in the first Brillouin zone via Eq.~(\ref{eq:lind}). Next, the Matsubara data are analytically continued to obtain $\chi^{R}(\omega,\mathbf{q})$. The theoretical $\chi^{R}(\omega,\mathbf{q})$ is then compared to the exact one as evaluated by Eq.~(\ref{eq:hard}).

The upper panels of Figure~\ref{fig:Y08} illustrate the momentum-resolve spectral functions. The panel (a) exhibits the exact spectrum, while the analytic continuation results obtained by the SPX method and MaxEnt method are shown in panels (b) and (c), respectively. Near the $\Gamma$ point, the exact spectrum manifests a single low-energy mode, which consists of low-energy excitations close to the Fermi surface. It is called the zero-sound mode. Along $\Gamma \to X$ or $\Gamma \to M$, the zero-sound mode will be broadened by Landau damping. We can see that both the SPX method and the MaxEnt method correctly capture the zero-sound mode and its broadening trend. Let us look at the $(\pi/2,0)$ point, which is indeed the midpoint of $\Gamma - X$. As is evident in Fig.~\ref{fig:Y08}(d), the simulated spectra are well consistent with the exact one. When going from $X$ to $M$, the exact spectrum acquires a simple structure. But it seems that both methods perform poorly here. For example, the exact spectrum for the $X$ point forms an asymmetric peak at approximately $\omega = 1.0$. However, the peak's center is shifted to $\omega = 0.8$ and the low-energy shoulder peak becomes more remarkable in the simulated spectra [see Fig.~\ref{fig:Y08}(e)]. As for the $M$ point, the exact spectrum consists of a relatively flat bump structure between $\omega = 1.0$ and $\omega = 2.0$. However, the simulated spectra consist of a spurious peak around $\omega = 1.3 \sim 1.4$ [see Fig.~\ref{fig:Y08}(f)]. In the previous study, Sch\"ott \emph{et al.} investigated the same model by using various analytic continuation methods~\cite{PhysRevB.94.245140}. They found that the MaxEnt, Pad\'{e}, NNT, and NNLS methods favor sharp peaks, instead of a broad flat feature for the $M$ point under intermediate noise ($\delta = 10^{-4}$). Only for very low noise ($\delta = 10^{-10}$), the flat bump structure can be recovered by these methods. Overall, the SOM method performs the best across various noise levels. But there are still small wiggles in the simulated spectrum.       

\section{Applications: Matrix-valued Green's functions\label{sec:application_matrix}}

In sections~\ref{sec:application_f} and \ref{sec:application_b}, we just examine the SPX method for analytical continuations of various fermionic and bosonic systems. Heretofore the correlation functions we treated are assumed from single-band models or diagonal components of matrix-valued Green's functions of multi-orbital models. However, quantum many-body computations do also provide off-diagonal Green's functions, such as in the DFT + DMFT context relevant for electronic structure calculations of strongly correlated materials~\cite{RevModPhys.78.865}. So, a question naturally arises. How about the SPX method for the matrix-valued Green's functions, especially for the off-diagonal components?

Since the spectral functions for the off-diagonal Green's functions might be negative and can not be interpreted as a probability distribution, the traditional MaxEnt method fails\cite{JARRELL1996133}. In the last decade, many efforts have been devoted to remedying this problem and improving the MaxEnt method. One possible way to overcome this limitation is to figure out an optimal basis in which the off-diagonal elements are eliminated~\cite{Tomczak_2007}. Only the diagonal elements are treated. Another way is to construct some auxiliary Green's functions with positive definite properties. Based on the analytic continuations of these auxiliary functions, the spectral functions of the off-diagonal components are derived. This is the so-called MaxEntAux method~\cite{PhysRevB.90.041110,PhysRevB.95.121104,PhysRevB.92.060509,yue2023maximum}. Recently, a notable idea has been proposed by Kraberger \emph{et al.}~\cite{PhysRevB.96.155128,KAUFMANN2023108519} They decomposed the spectral functions $A$ into $A^{+} - A^{-}$ and generalized the Shannon-Jaynes entropy $S[A]$ to the positive-negative entropy $S[A^+,A^-]$, such that the positive definite condition is satisfied and the traditional MaxEnt method works. Similarly, Sim and Han tried to extend the MaxEnt method by introducing the quantum relative entropy~\cite{PhysRevB.98.205102}. Their method is formulated in terms of matrix-valued function, and the Bayesian probabilistic interpretation is maintained just as the conventional MaxEnt method~\cite{JARRELL1996133}. In addition, Fei and Gull also extended their NAC method to support analytic continuations of matrix-valued Green's functions~\cite{PhysRevLett.126.056402,PhysRevB.104.165111}. Anyway, reliable methods for performing the analytic continuation of the whole Green's function matrix are highly desirable. In this section, we would like to benchmark the SPX method for two kinds of matrix-valued Green's functions. 

\subsection{Gaussian model\label{subsec:gauss_m}}

\begin{figure*}[ht]
\includegraphics[width=\textwidth]{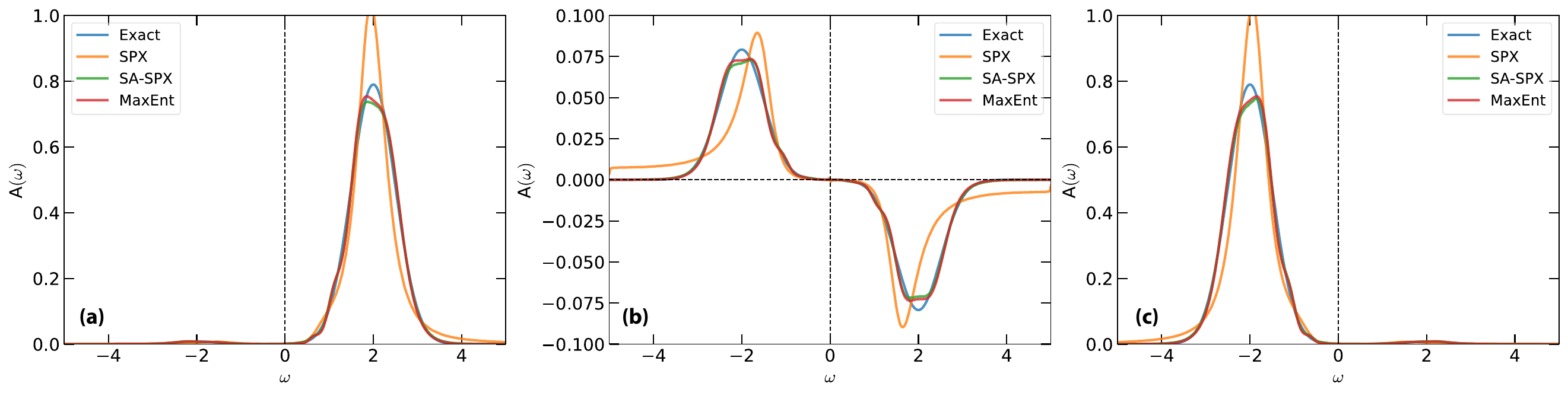}
\caption{Analytical continuations of matrix-valued Green's functions (M$_{12}$: Gaussian model). (a) Calculated and exact $A_{11}$. (b) Calculated and exact $A_{12}$. (c) Calculated and exact $A_{22}$. Note that in the SA-SPX method, the SPX method is combined with the self-adaptive sampling algorithm to refine the spectra. \label{fig:Z01}}
\end{figure*}

As the first example, we apply the SPX method to a simple two-band model. We adopt the procedure as described in Reference~[\onlinecite{PhysRevB.98.205102}] to construct the matrix-valued Green's function. At first, two spectral functions $A_{11}(\omega)$ and $A_{22}(\omega)$ are generated by using the Gaussian model [see Eq.~(\ref{eq:model01})]. The parameters are as follows: $N_{g,11} = N_{g,22} = 1$, $\epsilon_{11} = -\epsilon_{22} = 2.0$, $w_{11} = w_{22} = 1.0$, $\Gamma_{11} = \Gamma_{22} = 0.5$. Next, the two spectral functions form a $2 \times 2$ diagonal matrix, i.e., $A_{12}(\omega) = A_{21}(\omega) = 0.0$. This matrix is rotated by a rotation matrix $R$:
\begin{equation}
\label{eq:rotation}
R = 
\left[
\begin{array}{cc}
    \cos \theta & \sin \theta \\
    -\sin \theta & \cos \theta \\
\end{array}
\right]
\end{equation}
where $\theta = 0.1$ denotes the rotation angle. Clearly, for $\theta = 0.0$, the off-diagonal elements of the matrix-valued spectral functions are all zero. Third, the Matsubara Green's function is constructed via Eq.~(\ref{eq:spectral_f}). We choose the fermionic kernel [see Eq.~(\ref{eq:kernel_f})]. And random Gaussian noises are added to the Matsubara data to mimic a realistic QMC situation. Finally, we get a $2 \times 2$ matrix-valued Green's function.

Now we apply the MaxEnt method and the SPX method to treat this matrix function in an element-wise manner. For the diagonal elements ($\mathcal{G}_{11}$ and $\mathcal{G}_{22}$), they are treated as before since their spectral functions are all positive. While for the off-diagonal elements ($\mathcal{G}_{12}$ and $\mathcal{G}_{21}$), the positive-negative entropy approach suggested by Kraberger \emph{et al.}~\cite{PhysRevB.96.155128} is employed. As for the SPX method, the self-adaptive sampling algorithm is used to refine the calculated spectra. The number of self-consistent iterations is around 5~\cite{niter}. The number of poles for positive and negative parts is equal (i.e. $N^+_p = N^-_p = N_p / 2$). The analytic continuation results are visualized in Figure~\ref{fig:Z01}. We can see that with the help of the self-adaptive sampling algorithm, the spectral functions obtained by the SPX method agree quite well with those obtained by the MaxEnt method and the exact spectra, irrespective of the diagonal or off-diagonal elements. 

\subsection{Pole model\label{subsec:pole_m}}

\begin{figure*}[ht]
\includegraphics[width=\textwidth]{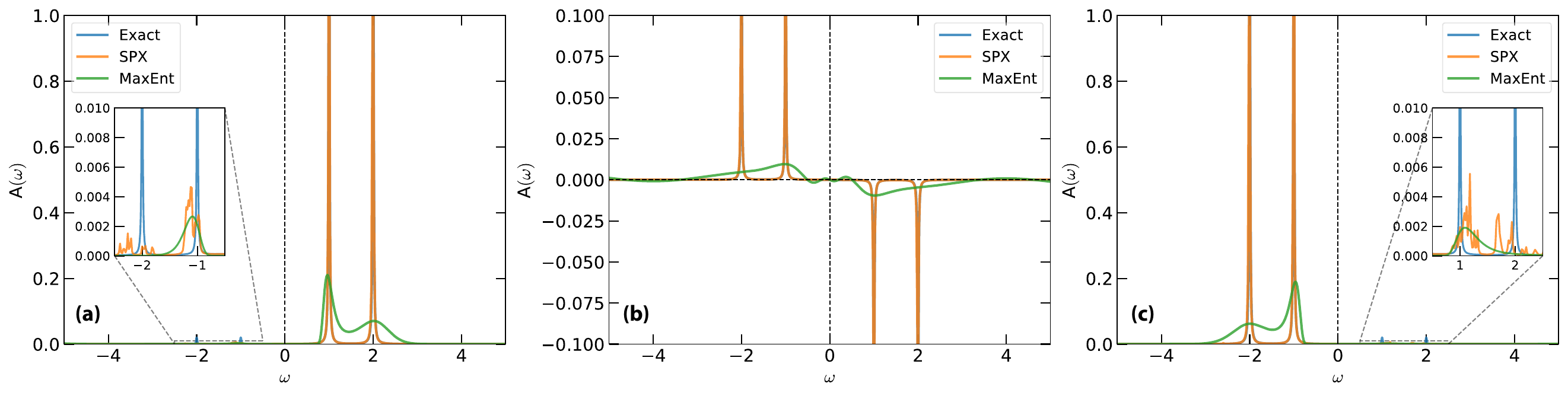}
\caption{Analytical continuations of matrix-valued Green's functions (M$_{13}$: Pole model). (a) Calculated and exact $A_{11}$. (b) Calculated and exact $A_{12}$. (c) Calculated and exact $A_{22}$. The insets in (a) and (c) show two small peaks that are easily overlooked. \label{fig:Z02}}
\end{figure*}

In the first example, we examine a condensed matter case in which the spectrum is usually broad and smooth. In this example, let us turn to a typical molecule case. We adopt a two-pole model to construct the spectral functions for diagonal elements [see Eq.~(\ref{eq:pole})]. The parameters are as follows: $N_{p}^{11} = N_{p}^{22} = 2$, $A^{11}_{1} = A^{11}_{2} = A^{22}_1 = A^{22}_2 = 0.5$, $P^{11}_1 = -P^{22}_1 = 2.0$, $P^{11}_2 = -P^{22}_2 = 1.0$. Next, the spectral functions are rotated to build the matrix-valued Green's function. The rotation matrix $R$ is given by Eq.~(\ref{eq:rotation}). The rotation angle $\theta$ is also 0.1, similar to the first example. 

As is evident in Figure~\ref{fig:Z02}, the SPX method works quite well. Not only the diagonal elements but also the off-diagonal elements are accurately resolved. Especially, the four trivial peaks in -2.0 and -1.0 for $A_{11}$ and in 2.0 and 1.0 for $A_{22}$ are also captured~\cite{trivial}. On the contrary, the MaxEnt method completely fails just as expected. For the diagonal elements, it tends to generate smooth peaks instead of sharp $\delta$-like peaks. For the off-diagonal elements, unphysical oscillations appear near the Fermi level. So, at least in this example, the SPX method is superior to the MaxEnt method.

\section{Discussions\label{sec:discuss}}

\subsection{Noisy data\label{subsec:noise}}

\begin{figure*}[ht]
\includegraphics[width=\textwidth]{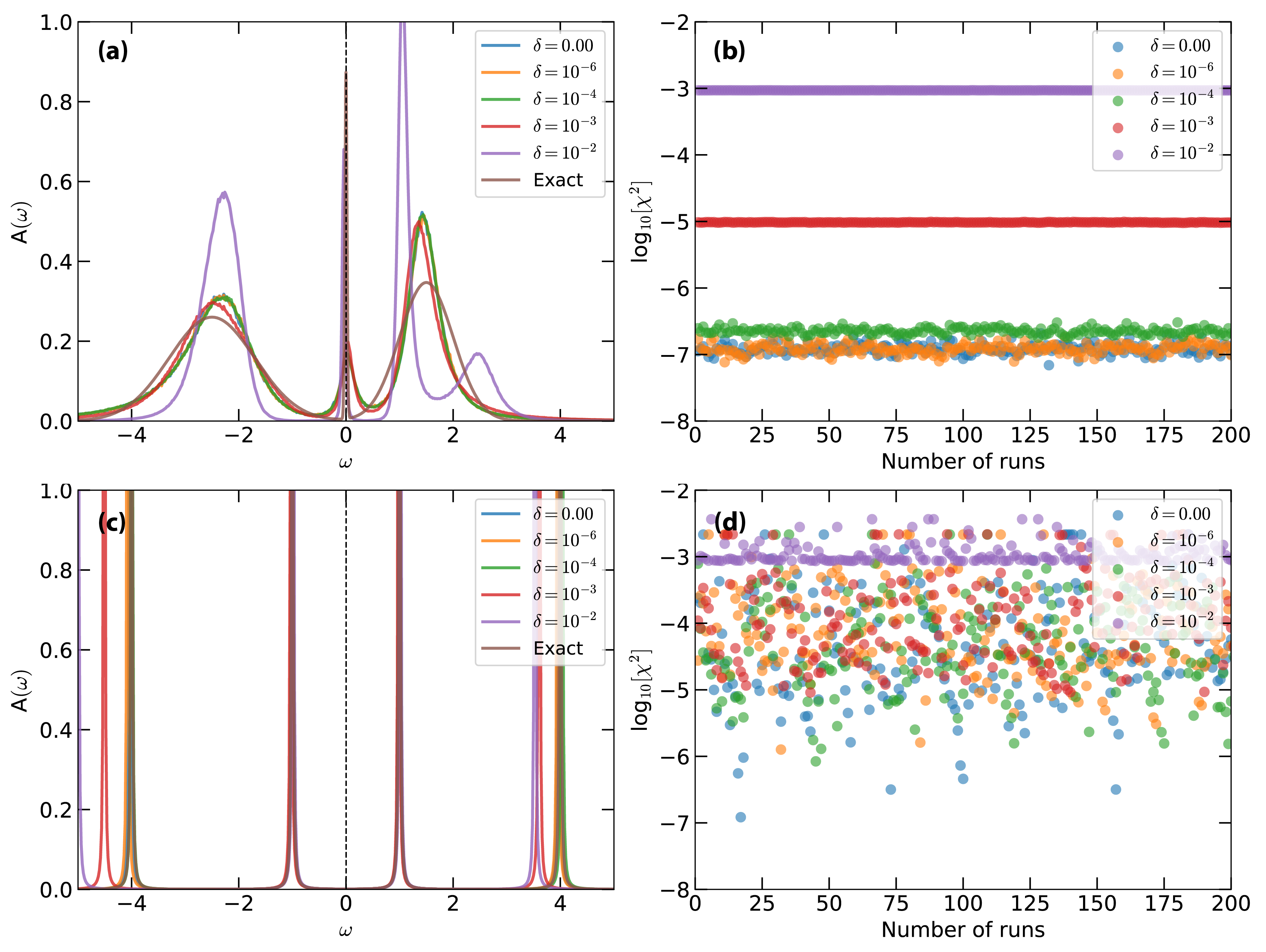}
\caption{Robustness of the SPX method with respect to noisy Matsubara data for fermionic correlators. Panels (a)-(b) are for the M$_{01}$ model (scenario iii, three Gaussian-like peaks), while panels (c)-(d) are for the M$_{02}$ model (scenario ii, four $\delta$-like peaks). (a) and (c) Synthetic and calculated spectral functions. (b) and (d) Noise-dependent goodness-of-fit functions $\chi^{2}$. Here, $\delta$ denotes the noise level. See Section~\ref{sec:setup} for more details. \label{fig:N03}}
\end{figure*}

\begin{figure}[ht]
\includegraphics[width=\columnwidth]{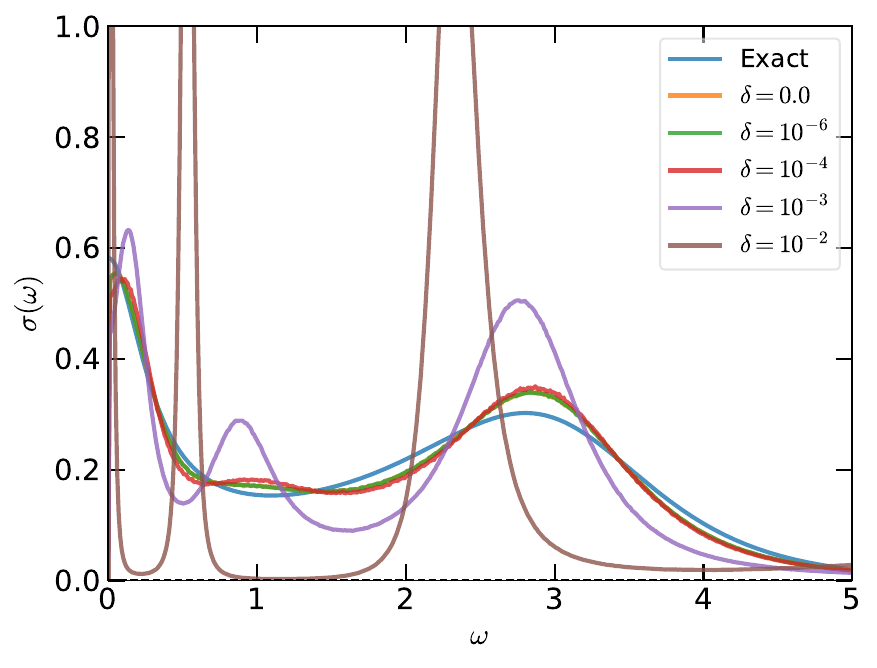}
\caption{Robustness of the SPX method with respect to noisy Matsubara data for bosonic correlators. The results for the M$_{10}$ model are shown. The size of input Matsubara data is fixed at 40. \label{fig:Y07N}}
\end{figure}

Matsubara data from realistic quantum many-body simulations are usually noisy~\cite{PhysRevB.84.075145,PhysRevB.97.205111,PhysRevB.96.035147}. The analytic continuation methods that based on the interpolation approach, such as the Pad\'{e}~\cite{PhysRevB.61.5147,PhysRevB.93.075104,Vidberg1977}, NAC~\cite{PhysRevLett.126.056402}, and Carath\'{e}odory~\cite{PhysRevB.104.165111} methods, are quite sensitive to the noise embedded in the input data. In the presence of moderate noise, these methods often suffer from unphysical oscillations or violate the causality of the spectral function.        

Here we would like to demonstrate the robustness of the SPX method with respect to noisy Matsubara data. As mentioned above, the $\delta$ parameter is used to control the magnitude of noise [see Eq.~(\ref{eq:noise})]. The larger the $\delta$ parameter is, the noisier the Matsubara data are. We regenerate the Matsubara Green's functions with various noise levels by using the workflow as introduced in Section~\ref{sec:setup}. Four noise levels, i.e., $\delta = 0.0$, $10^{-6}$, $10^{-4}$, $10^{-2}$, are considered. 

Let us treat the fermionic systems at first. The M$_{01}$ and M$_{02}$ models are selected as representative cases. Figure~\ref{fig:N03} shows the benchmark results. When the Matsubara data are clean ($\delta = 0.0$), almost all the features of the spectral function of the M$_{01}$ model are well resolved, and all the sharp peaks in the spectral function of the M$_{02}$ model are perfectly retrieved. When the noise level is small or moderate ($\delta = 10^{-6}$, $10^{-4}$, or $10^{-3}$), the analytic continuation results are almost the same with those for noiseless Matsubara data, and the goodness-of-fit function remains small. When the noise level is large ($\delta = 10^{-2}$), the performance of the SPX method is more or less affected. The goodness-of-fit function gets worse. For the M$_{01}$ model, the position and weight of the lower Hubbard band are retrieved. The quasiparticle resonance peak is well reproduced. However, the upper Hubbard band is split into two peaks, and two sizable gaps emerge between the quasiparticle resonance peak and the lower/upper Hubbard band. As for the M$_{02}$ model, the low-frequency peaks are well reproduced, but the positions of the high-frequency peaks are shifted slightly.

We have already revealed that the SPX method is quite robust for the fermionic Matsubara data. How about the bosonic functions? Now we would like to examine the robustness of the SPX method with respect to noisy Matsubara data for bosonic functions. The M$_{10}$ model (for the analytical continuation of the current-current correlation function) is selected as the test case. The analytic continuation results are shown in Fig.~\ref{fig:Y07N}. We find that the SPX method yields almost identical spectra for $0.0 \le \delta < 10^{-4}$. When $\delta = 10^{-3}$, the Drude peak is slightly depressed and a shoulder peak around $\omega = 0.9$ becomes prominent. Once $\delta = 10^{-2}$, the Drude peak becomes extremely narrow. The shoulder peak around $\omega = 0.9$ disappears, but a large spurious structure comes out at $\omega = 0.5$, which is probably due to the SPX method giving too much weight to the noise.

As a whole, it is suggested that the SPX method is noise-tolerant. It works quite well even when the noise level is relatively large.

\subsection{Incomplete data\label{subsec:incomplete}}

\begin{figure}[ht]
\includegraphics[width=\columnwidth]{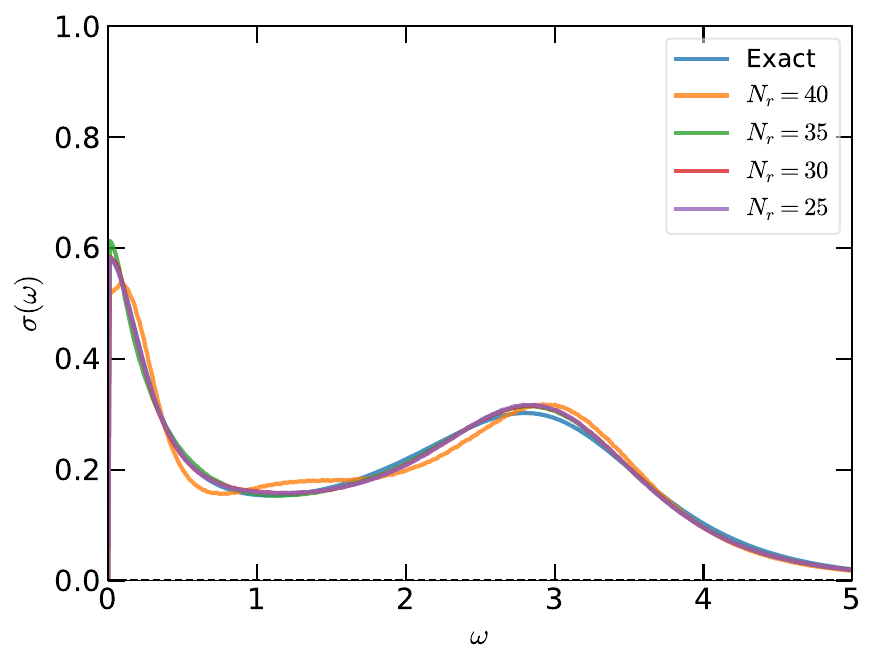}
\caption{Robustness of the SPX method with respect to incomplete data (M$_{10}$: Current-current correlation function). The raw input data contain 40 data points, and we try to remove some data from them randomly. The $N_r$ denotes the number of residual data points. \label{fig:Y07P}}
\end{figure}

Besides the noises, sometimes the input data could be broken. The purpose of this subsection is to discuss the robustness of the SPX method with respect to the incomplete Matsubara data. We take the M$_{10}$ model as the test case again. At first, we generate the input data using Eq.~(\ref{eq:current}) and Eq.~(\ref{eq:optic}). The size of full input data is $N_w = 40$. Since the first Matsubara frequency point $\omega_0$ is essential to realize the sum-rule for bosonic systems [see Eq.~(\ref{eq:sum_rule_b})], and the last Matsubara frequency point $\omega_{N_w - 1}$ is relevant with the high-frequency behaviors of the spectrum, so $\Pi(i\omega_0)$ and $\Pi(i\omega_{N_w - 1})$ are always kept. Then we try to pick and remove some data points randomly in the rest of the input data. We test four cases with $N_r = 25$, 30, 35, 40, where $N_r$ denotes the size of the residual data. 

The analytic continuation results are shown in Figure~\ref{fig:Y07P}. We find that the SPX method is insensitive to the completeness of the input data. Even when $N_r = 25$ (it implies that 37.5\% data points are unavailable), the calculated spectrum is still reasonable and exhibits little deviations when compared to the one obtained with full input data and the exact one.

\subsection{Comparison with the other methods\label{subsec:compare}}

\emph{Compared with stochastic analytic continuation}. As mentioned before, both the SPX method and the SAC methods~\cite{beach2004,PhysRevB.57.10287,PhysRevE.94.063308,PhysRevE.81.056701,PhysRevB.76.035115,PhysRevX.7.041072,PhysRevB.78.174429} are classified as the ASM approach~\cite{PhysRevB.78.174429,PhysRevB.101.085111,PhysRevB.102.035114,doi:10.1063/1.3185728}. So these methods share some common features. Now we will elaborate on their similarities and differences: (i) Both methods are based on the stochastic algorithms. They employ the Monte Carlo sampling algorithm to locate the global minimum of the loss function. (ii) In the SPX method, the correlator itself is parameterized by many poles in the complex plane. While in the SAC methods, the spectral function is parameterized by a large number of $\delta$ functions (or rectangle functions) in continuous frequency space. In other words, the SPX method tries to fit the $G(i\omega_n)$ data directly, while the SAC methods try to fit the $A(\omega)$. (iii) Both methods support imposing additional constraints to reduce the configuration space and accelerate the Monte Carlo sampling procedure.

\emph{Compared with pole fitting method}. Quite recently, Lin Lin \emph{et al.} proposed a three-pronged projection-estimation-semidefinite (PES) relaxation method~\cite{PhysRevB.107.075151} to perform analytic continuation of Green's functions. Their method adopts the pole representation of the Matsubara Green's function as well. At first glance, the SPX method and the PES method are quite similar. However, the key ideas of the two methods are completely different. Next, we would like to clarify this issue. 

The PES method consists of three steps: (i) The noisy Matsubara data are projected into the causal space. In this step, the Matsubara data are fitted by:
\begin{equation}
\label{eq:proj}
\mathcal{G}^{\text{proj}}(i\omega_n) = \sum^{M}_{m = 0} \frac{\mathcal{A}^{\text{proj}}_m}{i\omega_n - x_m}.
\end{equation}
Eq.~(\ref{eq:proj}) looks like Eq.~(\ref{eq:pole}). But, $x_m$ in Eq.~(\ref{eq:proj}) is a fine uniform grid on real axis:
\begin{equation}
x_m = \omega_{\text{min}} + \frac{m}{M}(\omega_{\text{max}} - \omega_{\text{min}}), \quad m = 0, 1, \cdots, M,
\end{equation} 
where $M + 1$ is the total number of grid points, $\omega_{\text{min}}$ and $\omega_{\text{max}}$ denote the left and right boundaries of the grid, respectively. The projection step can be considered as an analytic continuation method by itself, but its quality is greatly constrained by the resolution of the grid on the real axis. In other words, the number of grid points $M$ must be as large as possible in order to resolve the pole locations accurately. So, the objective of the projection step is to project the noise data into the causal space and filter the noise, instead of performing analytic continuation directly. (ii) The AAA algorithm is used to reduce the number of poles and estimate their locations~\cite{AAA,AAA_Lawson}. Since the AAA algorithm is sensitive to the noise level, it takes the projected Matsubara data as input. Note that the locations of poles given by the AAA algorithm are not accurate. They just serve as an initial guess. (iii) The SDR algorithm~\cite{PhysRevB.101.035143} is used to obtain an effective fitting of the Matsubara data in the form of Eq.~(\ref{eq:pole}). Here, the projected Matsubara data from the projection step, and the initial guess of the locations of the poles from the estimation step are taken as inputs. The SDR algorithm employs a bi-level optimization approach to locate the global minimum of loss function $\chi^2$.

Clearly, the PES method is a decisive approach. In the projection step, the locations of poles are fixed, and only the amplitudes of poles are optimized. This is a convex optimization problem and can be easily solved. In the next two steps, due to the limitations of the AAA and SDR algorithms, the number of poles must be relatively small. Therefore the PES method is useful for extracting the spectra of molecules and band structures in solids, which usually feature by multiple $\delta$-like peaks. But it is hard to resolve broad and smooth spectral functions if the poles reside in the real axis only~\cite{PhysRevB.107.075151}. Just as its name implies, the SPX method is a stochastic approach. Within the SPX method, both the amplitudes and locations of poles are optimized at the same time. This is a highly non-convex optimization problem. But, thanks to the simulated annealing algorithm~\cite{Marinari:1996dh,SA1983}, the optimization problem can be efficiently solved. There's no need to project the noisy data into causal space, make an initial guess for the amplitudes and locations of poles, and limit the number of poles. The complicated AAA and SDR algorithms are not required anymore. This is the major reason why the SPX method can be used to retrieve the spectral functions for both condensed matter cases and molecule cases.

\section{Concluding remarks\label{sec:summary}}

In the present work, we have developed a new stochastic approach, namely the SPX method, for analytic continuations of Matsubara Green's functions. In the SPX method, the Matsubara data are represented by a sequence of poles, and the amplitudes and locations are optimized by using the simulated annealing algorithm. Some representative examples, including the single-particle Green's functions, self-energy functions, two-particle correlation functions, and matrix-valued Green's functions, etc., are employed to benchmark the usefulness and robustness of this method. The calculated results are compared with the exact spectra if available and the ones obtained by the MaxEnt method. For most of the examples, the performance of the SPX method is comparable or superior to that of the MaxEnt method. Applications to the synthetic and realistic Matsubara data reveal that this new method could resolve not only low-frequency smooth peaks in condensed matter cases but also high-frequency sharp peaks in molecule cases. Thus, it provides a promising route to extract dynamical responses from imaginary frequency single-particle and two-particle correlation functions. 

The SPX method overcomes most of the shortcomings of the available analytic continuation methods and manifests competitive performance and applicability. The major advantages are summarized as follows: (1) It supports analytic continuations of fermionic correlators, bosonic correlators, and matrix-valued Green's functions. (2) The SPX method is rather robust to external noise. It works quite well at intermediate and low noise levels. It is also robust for incomplete Matsubara data. (3) The SPX method can recover broad peaks and multiple $\delta$-like peaks precisely. In other words, it supports analytic continuations for both condensed matter cases and molecule cases. (4) Combining with the well-designed constrained algorithm and self-adaptive sampling algorithm, the SPX method can resolve singular structures (such as sharp peaks and band edges) in the spectra. (5) Once the simulation is finished, the SPX method can yield approximate expressions for the Matsubara and retarded Green's functions. Thereby the tricky Kramers-Kronig transformation is avoided. 

In addition to solving the analytic continuation problems, the SPX method should have broad applications in the finite temperature quantum many-body simulations. For example, it could serve as a noise filter and generate inputs for the other noise-sensitive analytic continuation methods, such as the Pad\'{e}~\cite{Vidberg1977,PhysRevB.93.075104,PhysRevB.61.5147,PhysRevB.87.245135}, the NAC~\cite{PhysRevLett.126.056402} and Carath\'{e}odory methods~\cite{PhysRevB.104.165111}. It could generate training datasets for machine learning-assisted analytic continuation methods~\cite{PhysRevLett.124.056401,PhysRevB.98.245101,PhysRevB.105.075112,PhysRevResearch.4.043082,Yao_2022,Arsenault_2017}. It could be used to perform bath fitting in the dynamical mean-field theory~\cite{RevModPhys.68.13,PhysRevB.101.035143}, and evaluate the high-frequency asymptotic behavior of single-particle Green's function. It could provide a compact representation to store and manipulate the two-particle Green's function~\cite{PhysRevB.84.075145,PhysRevB.96.035147}. Thus, exploring further applications of the SPX method is highly demanded.

\begin{acknowledgments}
The authors thank Prof. Lei Wang for fruitful discussions. This work is supported by the Innovation Foundation of China Academy of Engineering Physics (No.~CX20200033), the National Natural Science Foundation of China (No.~12274380 and No.~11934020), the National Key Projects for Research and Development of China (No.~2021YFA1400400), the China Postdoctoral Science Foundation (No.~2021TQ0355), and the Special Research Assistant Program of Chinese Academy of Sciences.
\end{acknowledgments}

\bibliography{spx}

\begin{thebibliography}{101}
\bibitem [{nit()}]{niter}%
  \BibitemOpen
  \bibinfo {note} {We can use the spectra generated by the MaxEnt method as
  input to build the frequency grid for the poles. At this time, two or three
  iterations are probably enough.}\BibitemShut {Stop}%
\bibitem [{tri()}]{trivial}%
  \BibitemOpen
  \bibinfo {note} {With the help of the constrained sampling algorithm and the
  self-adaptive sampling algorithm, the positions of the four trivial peaks can
  be resolved accurately.}\BibitemShut {Stop}%
\end{thebibliography}

\end{document}